\documentclass{scrartcl}

\usepackage{amsfonts}
\usepackage{amsmath}
\usepackage{amssymb}
\usepackage{amsxtra}
\usepackage{epsfig}
\usepackage{amsfonts}
\usepackage{color}

\usepackage{algorithmicx}
\usepackage{algorithm}
\usepackage{algpascal}

\usepackage{color}

\usepackage{setspace}

\begin{document}

\title{
Generalization of Kirchhoff reflectivity
to go beyond modelling and inversion of first-order reflection data -
A review of the theory.
}

\author{
J\'er\'emie Messud\\
\textit{CGG, 27 avenue Carnot, 91341 Massy Cedex, FRANCE}
}

\maketitle

\begin{abstract}

We remind and emphasize the connections and differences
between Kirchhoff and Born modelling.
We discuss how they lead to a general expression
for the conversion of a velocity model perturbation into
a reflectivity through the ``generalized reflectivity" concept.
We seize the opportunity to clarify some aspects related to 
possibly non-smooth propagating media
and the linearity approximation on reflectors.

The generalized reflectivity offers opportunities:
\begin{itemize}
\item
On FWI approaches that include a reflectivity or least squares migration approaches that can be based on Kirchhoff or Born modelling:
to rigorously convert the reflectivity into a velocity perturbation.
\item
In the framework of traditional Kirchhoff modelling scheme:
to model first-order effects that go beyond first-order reflections (like first-order diffractions).
\item
In the framework of traditional Kirchhoff inversion or true amplitude migration,
i.e. for the interpretation of seismic-migrated images:
to give a basis to interpret by AVA (amplitude versus angle)
more information than the amplitudes associated to first-order reflections,
for instance the amplitudes of first-order diffractors.
Also, it would theoretically allow us to go beyond AVA analysis,
inverting for the whole seismic image amplitude information (not only amplitude information at peaks)
to recover the related velocity model perturbation.
This is discussed formally in this article.
\end{itemize}

\end{abstract}

\section{Introduction.}
\label{sec:intro}

The aim of seismic imaging \cite{Cla85} is to characterize the geological structures of the subsurface
from the analysis of seismic waves \cite{Aki80,Cha04}.
A central component of seismic imaging is the scale separation,
i.e. the separation of a smooth background velocity containing the long wavelength components
of the true subsurface velocity model from the short wavelength components \cite{Cla85,Tar05}.
This separation is justified regarding the physical behavior of these two components
with respect to band limited data (typically 3 to 80 Hz)
\cite{Cla85,Tar05,Vir15}:
the background velocity exhibits a strongly non-linear behavior with respect to the data,
affecting the kinematics of the seismic events,
while the short wavelength components have a much more linear behavior,
affecting mostly the amplitudes of the events.
As a consequence, recovery of the background velocity and of the short wavelength components
is usually done sequentially \cite{Lai83,Tar84}.
The first step is to compute the background velocity, typically
by non-linear tomographic methods \cite{Luo91}.
The second step is to compute the short wavelength components
through a linear inversion process, considering first-order scattered events (reflections and diffractions) \cite{Tar05,Cla71,Ble87}, called seismic migration or imaging.
There are two ways in seismic migration of linearly representing the short wavelength components of the velocity model:
\begin{itemize}
\item
Using the Born approximation \cite{Bey85,Bey86,Lam92,Ble}, based on a velocity model perturbation.
\item
Using the Kirchhoff approximation \cite{Cla85,Ble,Sto02,Kro98,Bra03},
where the short wavelength components are represented through a reflectivity distribution,
i.e. a volumetric distribution of reflection coefficients.
Inspired by the pioneering work of \cite{Bey85,Bey86},
Bleistein's groundbreaking work \cite{Ble87,Ble} fundamentally establishes the reflectivity
and shows how it can, at a later stage, be converted into material properties of the
subsurface through an additional inversion process
like AVA (amplitude versus angle) analysis \cite{Ble87,Rus88}.
\item
Both ways can be used in a least square version of migration \cite{Sym16,Kro14}. 
\end{itemize}

Full waveform inversion (FWI) \cite{Tar05,Vir09} is another approach for characterizing the subsurface velocity.
Its ultimate aim is to invert band-limited seismic data non-linearly for the full range of wavelength components of the velocity model.
In common FWI applications, a local optimization scheme is used
(each iteration being related to a linearization, i.e. the Born approximation) \cite{Tar05},
so that an initial velocity model that is sufficiently good kinematically is needed
to avoid local minima.
Then, the non-linearity is sufficiently weak and FWI can invert for long wavelength components.
A reflectivity can be introduced within FWI to model first-order reflections
but it must be converted into a velocity perturbation at each iteration
for the velocity update \cite{Cla85,Ber82,Xu12}.

Those two representations of the short wavelength 
components of the velocity model, i.e. model perturbation (Born) versus reflectivity (Kirchhoff),
are commonly used.
They are both based on a linearization, but each offers some specifics.
For instance, Born approximation allows for modelling of first-order reflections on weak discontinuities
and first-order diffractions,
whereas Kirchhoff approximation allows for modelling of first-order reflections possibly on stronger discontinuities and postcritical reflections.
The connections and differences between the two have been studied from different points of view, 
see e.g. \cite{Ble,Urs97,Ale02,Bey94}.
In this paper, we propose to emphasize those connections and differences in a refreshing way.

First, we briefly recall the chain of approximations leading to Kirchhoff modelling equations.
We seize the opportunity to clarify some aspects related to Kirchhoff modelling,
concerning possibly non-smooth propagating media 
and the linearity approximation on reflectors.

Then, we detail how Kirchhoff and Born modelling can lead to very similar expressions
and how we can derive a general expression
for the conversion from velocity model perturbation into
reflectivity (and conversely) through a ``generalized reflectivity" concept.
We propose a different demonstration than the one existing in the literature \cite{Urs97}.
We then point out, from a formal point of view,
the strengths and weaknesses of the Kirchhoff and Born modelling schemes.

The generalized reflectivity offers opportunities:
\begin{itemize}
\item
In the framework of traditional Kirchhoff modelling scheme:
to model first-order effects that go beyond first-order reflections (like first-order diffractions).
\item
On FWI approaches that include a reflectivity \cite{Xu12} or least squares migration approaches that can be based on Kirchhoff or Born modelling \cite{Sym16,Kro14}:
to rigorously convert the reflectivity into a velocity perturbation.
\item
In the framework of traditional Kirchhoff inversion or true amplitude migration,
i.e. for the interpretation of seismic-migrated images:
to interpret by AVA
more information than the amplitudes associated to first-order reflections,
for instance the amplitudes of first-order diffractors.
Also, it would theoretically allow us to go beyond AVA analysis,
inverting for the whole seismic image amplitude information (not only amplitude information at peaks)
to recover the related velocity model perturbation.
This is discussed formally in this article.
\end{itemize}

\section{Kirchhoff modelling and inversion.}
\label{sec:Kirr1}

In the following, $t$ represents time, $\mathbf{r}=(x,y,z)$, position in the subsurface,
$\mathbf{r}_s$, position of an impulsive source of signature $s(t)$
and $\mathbf{r}_r$, the receiver positions.
Our time-direction Fourier transform convention is
$
A(\omega)= \int_{-\infty}^{+\infty} dt
\hspace{0.5mm} e^{-i\omega t} a(t)
$.
We use capital letters for the Fourier transform result.

Seismic waves are frequently modelled assuming a constant density acoustic approximation,
i.e. using the scalar wave equation where the subsurface model is parameterized by the velocity. 
The subsurface wavefield $p(\mathbf{r}_s,\mathbf{r},t)$ generated by a point source at $\mathbf{r}_s$
then obeys
\begin{eqnarray}
\text{For } \mathbf{r}_s\in \Re e^3,
\forall \mathbf{r}\in \Re e^3:
&&
\Big[ \frac{1}{c^2(\mathbf{r})}\frac{\partial^2}{\partial t^2} - \Delta \Big] p(\mathbf{r}_s,\mathbf{r},t)
=
\delta(\mathbf{r}-\mathbf{r}_s)
s(t)
\nonumber\\
&&
p(\mathbf{r}_s,\mathbf{r},t)=0
\quad\text{and}\quad
\frac{\partial}{\partial t}
p(\mathbf{r}_s,\mathbf{r},t)=0
\quad\text{for}\quad
t \le 0
.
\label{eq:wave1}
\end{eqnarray}
$c$ is the velocity of the subsurface.
$p$ is assumed to satisfy proper boundary conditions (free surface and Sommerfeld radiation condition).

\subsection{Kirchhoff modelling.}
\label{sec:Kirr1_1bis}


Reflectors (or ``smooth physical interfaces") \cite{Ble} are defined by discontinuities
in the subsurface model $c$ that generate reflections.
Reflections are defined within 0-order geometrical optics (0-g.o.) or high-frequency approximation \cite{Cer,Kra90}
by the events that satisfy the Snell-Descartes law,
which imposes a particular direction to a reflected ray according
to the direction of the corresponding incident ray \cite{Ble,Cer,Kra90}.
(Contrariwise diffraction events do not satisfy the Snell-Descartes law:
diffracted rays radiate in all directions.)

\begin{figure}[ht]
\begin{center}
\includegraphics[angle=0,width=0.9\linewidth]{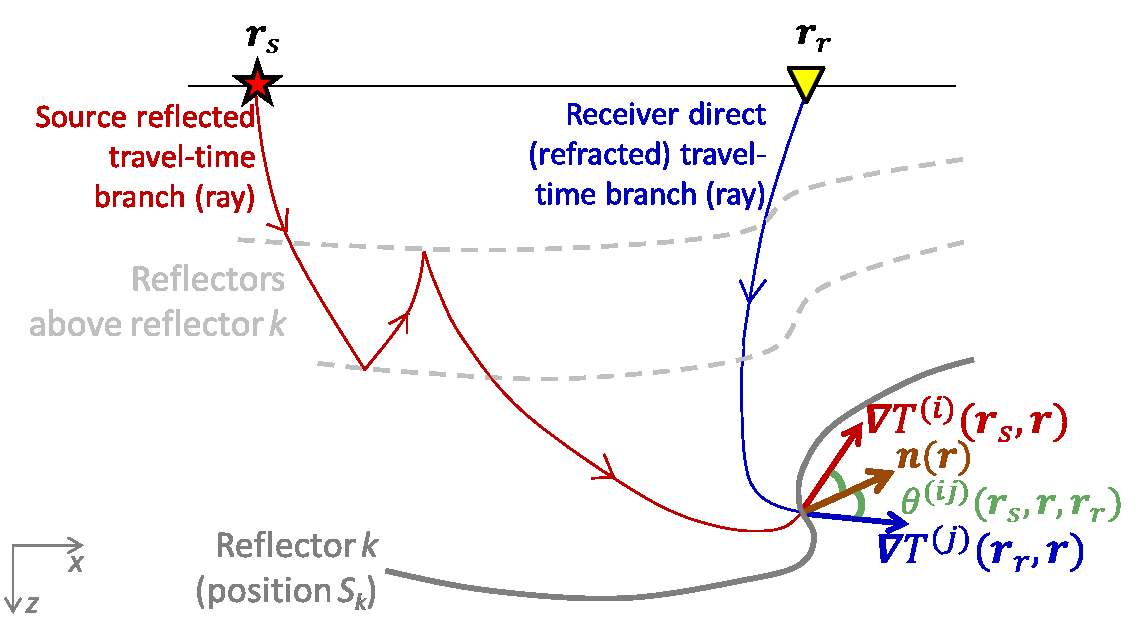}
\caption{
Source and receiver travel-time branches in a configuration where they are related by a reflection
from above on a given reflector $k$.
\label{fig:branches}
}
\end{center}
\end{figure}

In the following, we consider a subsurface composed of infinitely spread (or ``extended") and sufficiently separated reflectors
(in a sense that will be clarified later),
with a sufficiently smooth velocity between reflectors
from the 0-g.o. point of view.
$S_k$ denotes the position of one reflector surface in the subsurface.
``Above $S_k$" means in the ``incident" medium by slight abuse of language, see Fig. \ref{fig:branches}.


The Green function $g(\mathbf{r}_s,\mathbf{r},t)$ of the subsurface satisfies eq. (\ref{eq:wave1}) with $s(t)=\delta(t)$.
It is decomposed above $S_k$ into:
\begin{itemize}
\item
An ``incident" field $g_{inc}$ that is generated by the source and does not interact
with $S_k$ and the medium below $S_k$;
in other terms it satisfies eq. (\ref{eq:wave1}) above  $S_k$ with $s(t)=\delta(t)$
and radiation (or ``absorbing" boundary) conditions on  $S_k$.
\item
A field $g_{ref}=g-g_{inc}$ that represents what remains, i.e. reflections generated on $S_k$ and events generated below $S_k$
that ``come back" into the medium above $S_k$.
They are described through a boundary condition on $S_k$.
\end{itemize}

We briefly remind the main steps that lead to the Kirchhoff modelling approximation;
details are given in Appendix \ref{app:Kirr1_1}.

Firstly,
the 0-g.o. approximation \cite{Cer,Kra90} for the Green function $G_{inc}$
makes it possible to separate $G_{inc}$ into the contributions related to each of the
travel-time branches (or ray paths) that reach $S_k$ from the source or receiver sides.
\begin{itemize}
\item
The source-side Green function is
\footnote{
\label{foot:as}
When we use g.o.,
the Green functions implicitly 
become ``analytical signals",
obtained from the real signal by equating to $0$ the negative frequencies,
to be consistent with possibly complex reflection c\oe fficients \cite{Cer,Kra90}.
In other words, all equations that imply the 0-g.o.
approximation are implicitly defined for $\omega\ge 0$ only.
The real signal is recovered by ``symmetrizing" the negative frequencies.
}:
\begin{eqnarray}
\forall \mathbf{r}\in S_k:
\quad
G_{inc}(\mathbf{r}_s,\mathbf{r},\omega) 
&\approx&
\sum_{j\ge 1}
G_{inc}^{(i)}(\mathbf{r}_s,\mathbf{r},\omega) 
\nonumber\\
G_{inc}^{(i)}(\mathbf{r}_s,\mathbf{r},\omega) 
&=&
A^{(i)}(\mathbf{r}_s,\mathbf{r}) e^{-i\omega T^{(i)}(\mathbf{r}_s,\mathbf{r})}
,
\label{eq:Cler_0go-2bis}
\end{eqnarray}
where $i$ denotes the travel-time branch numbers.
$N(\mathbf{r}_s)$ denotes the number of direct travel-time branches (i.e. non-reflected, or refracted due to velocity inhomogeneities) and $i\le N(\mathbf{r}_s)$ refers to these arrivals.
$i>N(\mathbf{r}_s)$ refers to travel-time branches reflected (once or multiple times)
within the medium above $S_k$ but not on $S_k$
(remembering that $G_{inc}$ denotes the field that does not interact with $S_k$ and the medium below).
The wavefield on $S_k$ related to a travel-time branch $(i)$ is parameterized by
an amplitude $A^{(i)}$ and travel-time $T^{(i)}$ 
that satisfy respectively the transport and eikonal equations \cite{Cer,Kra90}.
$\nabla_\mathbf{r} T^{(i)}(\mathbf{r}_s,\mathbf{r}\in S_k)$ defines the direction
of the $i^{th}$ source-side travel-time branch ``ray" on the reflector.
\item
The receiver-side Green function
is defined in a similar way that in eq. (\ref{eq:Cler_0go-2bis}),
with $\mathbf{r}_s \rightarrow \mathbf{r}_r$ and $i\rightarrow j$ in the notations above.
\end{itemize}
Fig. \ref{fig:branches} gives an illustration, with $i>N(\mathbf{r}_s)$ for the source travel-time branch
and $j\le N(\mathbf{r}_r)$ for the receiver travel-time branch.
An important quantity is the ``incidence angle" $\theta_{inc}^{(i)}(\mathbf{r}_s,\mathbf{r})$ that is the acute angle at position $\mathbf{r}\in S_k$ between 
$\nabla_\mathbf{r} T^{(i)}$
and the unit vector $\mathbf{n}$ normal to
$S_k$ that points ``downward" (well defined for smooth $S_k$ only).
$\theta^{(ij)}(\mathbf{r}_s,\mathbf{r},\mathbf{r}_r)$ denotes half the angle at reflector positions
between the $i^{th}$ source travel-time branch ray
and the $j^{th}$ receiver travel-time branch ray.
For ``specular" source and receiver ray pairs, 
i.e. that satisfy the Snell-Descartes law for reflections \cite{Ble,Urs97},
we have
$\theta_{inc}^{(i)}(\mathbf{r}_s,\mathbf{r}) = \theta^{(ij)}(\mathbf{r}_s,\mathbf{r},\mathbf{r}_r)$, see Fig \ref{fig:branches}.

Secondly, the 0-g.o. (possibly complex) reflection c\oe fficient on $S_k$ is introduced \cite{Cer,Kra90,Ble,Ber82}
\begin{eqnarray}
&&
\hspace{-0.7cm}
\forall \mathbf{r}\in S_k:
\nonumber\\
&&
R^{(i)}(\mathbf{r}_s,\mathbf{r})
=
R(\mathbf{r},\theta_{inc}^{(i)}(\mathbf{r}_s,\mathbf{r}))
=
\lim_{\epsilon\rightarrow 0^+}
\frac{
c_{+}(\mathbf{r})
\cos(\theta_{inc}^{(i)}(\mathbf{r}_s,\mathbf{r}))
-
\sqrt{
c_{-}^2(\mathbf{r})
-\sin^2(\theta_{inc}^{(i)}(\mathbf{r}_s,\mathbf{r}))
c_{+}^2(\mathbf{r})
}
}
{
c_{+}(\mathbf{r})
\cos(\theta_{inc}^{(i)}(\mathbf{r}_s,\mathbf{r}))
+
\sqrt{
c_{-}^2(\mathbf{r})
-\sin^2(\theta_{inc}^{(i)}(\mathbf{r}_s,\mathbf{r}))
c_{+}^2(\mathbf{r})
}
}
\nonumber\\
&&
c_{+}(\mathbf{r})=\lim_{\epsilon\rightarrow 0^+}c(\mathbf{r} + \epsilon \mathbf{n}(\mathbf{r})),
\quad
c_{-}(\mathbf{r})=\lim_{\epsilon\rightarrow 0^+}c(\mathbf{r} - \epsilon \mathbf{n}(\mathbf{r}))
\label{eq:ref72}
.
\end{eqnarray}

Kirchhoff approximation considers only single events reflected from above on $S_k$,
i.e. source and receiver travel-time branches coupled with a single reflection
from above on reflector $k$.
This allows to relate linearly, within 0-g.o., the Green function $G_{ref}$ at the earth's surface to the reflection c\oe fficients on $S_k$ and to the Green function $G_{inc}$
in the medium above $S_k$.

$P_{ref}(\mathbf{r}_s,\mathbf{r}_r,\omega)=S(\omega)G_{ref}(\mathbf{r}_s,\mathbf{r}_r,\omega)$
denotes the total reflected wavefield measured at the earth's surface.
$P_{ref,k}^{(ij)}(\mathbf{r}_s,\mathbf{r}_r,\omega)$ represents the contribution 
of the $i^{th}$ ``source" travel-time branch, coupled with a single reflection
from above on reflector $k$ to the $j^{th}$ ``receiver" travel-time branch.
Note that in traditional Kirchhoff modelling approximation,
only the 
contributions 
related to direct (or refracted) source and receiver travel-time branches are considered,
i.e. $i\in[1,N(\mathbf{r}_s)]$ and $j\in[1,N(\mathbf{r}_r)]$.
Appendix \ref{app:Kirr1_main_steps} discusses the generalization to $i>N(\mathbf{r}_s)$ and $j>N(\mathbf{r}_r)$.

Appendix \ref{app:Kirr1_1} gives details on the computation
that leads to the Kirchhoff modelling approximation considering one reflector, $S_k$.
Then, suppose the subsurface reflectors are in a
configuration where they are separable almost everywhere,
i.e. a not too dense configuration in a sense that will be clarified in \S  \ref{sec:Kirr2_1_3}.
Performing the 
linearity approximation on reflectors
detailed in Appendix \ref{app:Kirr1_2}, we sum each reflectors $S_{k\ge 1}$ contribution and obtain the traditional Kirchhoff modelling approximation equation \cite{Ble}:
\begin{eqnarray}
&&
P_{ref}(\mathbf{r}_s,\mathbf{r}_r,\omega)
\approx
\sum_{k\ge 1}
\sum_{i=1}^{N(\mathbf{r}_s)}
\sum_{j=1}^{N(\mathbf{r}_r)}
P_{ref,k}^{(ij)}(\mathbf{r}_s,\mathbf{r}_r,\omega)
\nonumber\\
&&
P_{ref,k}^{(ij)}(\mathbf{r}_s,\mathbf{r}_r,\omega)
=
\int_{S_k}
d\mathbf{r} \hspace{0.5mm}
R(\mathbf{r},\theta^{(ij)}(\mathbf{r}_s,\mathbf{r},\mathbf{r}_r))
\frac{2\cos\big( \theta^{(ij)}(\mathbf{r}_s,\mathbf{r},\mathbf{r}_r) \big)}
{c(\mathbf{r})}
L_{inc}^{(ij)}(\mathbf{r}_s,\mathbf{r}_r,\mathbf{r},\omega)
\nonumber\\
&&
L_{inc}^{(ij)}(\mathbf{r}_s,\mathbf{r}_r,\mathbf{r},\omega)
=
i \omega S(\omega) 
G_{inc}^{(i)}(\mathbf{r}_s,\mathbf{r},\omega)
G_{inc}^{(j)}(\mathbf{r},\mathbf{r}_r,\omega)
.
\label{eq:Kir6-kbisbis}
\end{eqnarray}
All the performed approximations to obtain eq. (\ref{eq:Kir6-kbisbis}) are valid for sufficiently high-frequencies \cite{Ble}, except
the linearity approximation on reflectors that
is physically valid for not too large velocity contrasts on the reflectors
(even if eq. (\ref{eq:Kir6-kbisbis}) is mathematically well defined for large contrasts).
Subtleties about the linearity approximation on reflectors
are discussed in Appendix \ref{app:Kirr1_2},
related to non-direct travel-time branches and incident Green functions that can be different for different reflectors.
Remind we also did a not-too-dense reflectors configuration hypothesis.

Interestingly, the stationary phase approximation principle described for instance in Ref. \cite{Ble} allows the following replacement in eq. (\ref{eq:Kir6-kbisbis}), even for non specular ray pair \cite{Ble,Ble87,Urs97}:
\begin{eqnarray}
\label{eq:theta_replac_bis}
\forall \mathbf{r}\in S_k:
\hspace{2.5cm}
\theta_{inc}^{(i)}(\mathbf{r}_s,\mathbf{r})
\quad
&\Leftrightarrow&
\quad
\theta^{(ij)}(\mathbf{r}_s,\mathbf{r},\mathbf{r}_r)
.
\nonumber
\end{eqnarray}

The $G_{inc}^{(i)}$ were defined using 0-g.o., eq. (\ref{eq:Cler_0go-2bis}), for the demonstration of the Kirchhoff modelling equation.
But, as we factorized them in the final result, eq. (\ref{eq:Kir6-kbisbis}), a wave propagation scheme can also be used for their computation.
Ultimately, computing the $G_{inc}^{(i)}$ should involve the true subsurface velocity $c$, considering only a direct travel-time branch.
This is not always easy as $c$ can contain discontinuities. For instance within a 0-g.o. propagation this
would imply resolving boundary conditions through each discontinuity in $c$.
Within a wave propagation this would imply ``muting" all reflections
or the use of one-way propagators \cite{Cla85,Ber82}.
To avoid the need for this, it is common practice to introduce in the modelling a smooth velocity
$c\rightarrow c_{inc}$ that best reproduces travel-times and amplitudes of a wavefield generated at the earth's surface and measured at the reflector positions.
A smooth velocity $c_{inc}$ that meets as well as possible those criteria
can be defined through tomography \cite{Woo08,Lam08}.

However, if a strong reflector (i.e. a large velocity contrast such as a salt dome)
is present in the true subsurface, i.e. in $c$,
the use of a unique smooth velocity $c_{inc}$ will not be able to reproduce
good amplitudes at positions below the reflector.
A solution might involve considering two different smooth velocities:
$c_{inc}^{above}$ for propagations related to events occurring above the large contrast
and $c_{inc}^{below}$ for propagations related to events occurring below the large contrast,
i.e. for propagations from the surface through this large contrast.
This would permit more freedom in the modelling
to better reproduce the phase and amplitude below the reflector
(see for instance \cite{Yar13}).
The considerations of Appendix \ref{app:Kirr1_2} allow to fundamentally understand that
this remains in the spirit of the most general form of the linearity approximation on reflectors,
where the incident Green functions can be different for different reflectors.

\subsection{Reflectivity distribution, Kirchhoff inversion and interpretation.}
\label{sec:Kirr2_1_3}

We convert the surface integral in eq. (\ref{eq:Kir6-kbisbis}) into a volume integral 
to introduce the reflectivity $\hat{R}$,
i.e. a volumetric distribution of the reflection c\oe fficients.
We use the ``singular function of the reflector's surface",
i.e. the Dirac delta distribution $\delta_{S_k}(\mathbf{r})$ that spikes on $S_k$
\begin{eqnarray}
\text{For any volume $V'$ that contains $S_k$}
\quad:\quad
\int_{S_k} d\mathbf{r} \hspace{0.5mm} A(\mathbf{r})
=\int_{V'} d\mathbf{r} \hspace{0.5mm} A'(\mathbf{r}) \delta_{S_k}(\mathbf{r})
,
\label{eq:sing_ft}
\end{eqnarray}
where $A'(\mathbf{r}\in V')$ is any well-behaved extension of function $A$ defined only for $\mathbf{r}\in S_k$
in the whole volume $V'$.
%
%
If $g_k(\mathbf{r})=0$ is an equation that defines the position of the surface $S_k$,
its singular function is defined by \cite{Ble}
\begin{eqnarray}
\delta_{S_k}(\mathbf{r}) = |\nabla g_k(\mathbf{r})| \delta(g_k(\mathbf{r}))
.
\label{eq:sing_ft}
\end{eqnarray}
%
%
We obtain (choosing $V'$ to be the whole space under the earth's surface located at $z=0$)
\footnote{
$R$ is extended in the whole volume
through eq. (\ref{eq:ref72}) where $\mathbf{n}$ is continuously extended in-between reflector positions
(for instance in the average direction of source and receivers rays - then $R$ is different from $0$ only along reflectors).
} \cite{Ble}
%
\begin{eqnarray}
&&
P_{ref}(\mathbf{r}_s,\mathbf{r}_r,\omega)
\approx
\sum_{i=1}^{N(\mathbf{r}_s)}
\sum_{j=1}^{N(\mathbf{r}_r)}
\int_{z\ge 0}
d\mathbf{r} \hspace{0.5mm}
\hat{R}(\mathbf{r},\theta^{(ij)}(\mathbf{r}_s,\mathbf{r},\mathbf{r}_r))
L_{inc}^{(ij)}(\mathbf{r}_s,\mathbf{r}_r,\mathbf{r},\omega)
\nonumber\\
&&
\hat{R}(\mathbf{r},\theta^{(ij)}(\mathbf{r}_s,\mathbf{r},\mathbf{r}_r))
=
\sum_{k\ge 1} R(\mathbf{r},\theta^{(ij)}(\mathbf{r}_s,\mathbf{r},\mathbf{r}_r))
\frac{2\cos\big( \theta^{(ij)}(\mathbf{r}_s,\mathbf{r},\mathbf{r}_r)  \big)}{c_{inc}(\mathbf{r})}
\delta_{S_k}(\mathbf{r})
.
\label{eq:Kir_1_ref_final}
\end{eqnarray}

Eq. (\ref{eq:Kir_1_ref_final}) represents the Kirchhoff volumetric modelling equation \cite{Ble}.
It is based on the reflectivity $\hat{R}$, that represents a volumetric distribution
that ``points" on reflectors through $\delta_{S_k}$,
and contains information on the reflection c\oe fficients through $R$.

The reflectivity concept becomes interesting in the context of Kirchhoff inversion.
Suppose we recorded seismic data $P$
at the earth's surface, pre-processed
to retain only first-order reflections
(especially multiple reflections
having been filtered out).
Suppose we also produced a smooth subsurface model $c_{inc}$
that allows computing $L_{inc}^{(ij)}$.
One can then invert the linear equation (\ref{eq:Kir_1_ref_final}) to recover the reflectivity $\hat{R}$.

Let us consider only given travel-time branches, i.e. given $i$ and $j$ values
(for instance the ones related to the shortest travel-times).
Using eq. (\ref{eq:Kir_1_ref_final}), valid for reflections,
we return to $\theta_{inc}^{(i)}$ ($\Leftrightarrow$ $\theta^{(ij)}$) for the practical purpose
of removing the $\mathbf{r}_r$ dependency of the reflectivity and allow inversions per full shot,
$\hat{R}(\mathbf{r}_s,\mathbf{r})\leftarrow\hat{R}(\mathbf{r},\theta_{inc}^{(i)}(\mathbf{r}_s,\mathbf{r}))$ $\Leftrightarrow$ $\hat{R}(\mathbf{r},\theta^{(ij)}(\mathbf{r}_s,\mathbf{r},\mathbf{r}_r))$.
One then obtains the following linear inversion for each shot (i.e. each $\mathbf{r}_s$)
\begin{eqnarray}
\forall \mathbf{r}_s:
\quad
\hat{R}_{inv}(\mathbf{r}_s,\mathbf{r})
=
\arg\min_{\hat{R}(\mathbf{r}_s,\mathbf{r})}
\int
d\omega 
\int
d\mathbf{r}_r \hspace{0.5mm}
\Big|
P(\mathbf{r}_s,\mathbf{r}_r,\omega)
-
\int_{z\ge 0}
d\mathbf{r} \hspace{0.5mm}
\hat{R}(\mathbf{r}_s,\mathbf{r})
L_{inc}^{(ij)}(\mathbf{r}_s,\mathbf{r}_r,\mathbf{r},\omega)
\Big|^2
.
\label{eq:Kir_inv_1}
\end{eqnarray}
This is called least-squares Kirchhoff inversion or true amplitude migration 
by shots \cite{Cla85,Tar05,Ble87}.
The reflectivity $\hat{R}(\mathbf{r}_s,\mathbf{r})$ is rigorously speaking a singular distribution, that only lives within an integral, but the result of the band-limited (limited $\omega$ range) and aperture-limited (limited $\mathbf{r}_r$ range)
inversion (\ref{eq:Kir_inv_1}) gives an estimate $\hat{R}_{inv}(\mathbf{r}_s,\mathbf{r})$ of the reflectivity ``convolved" with a filter, that can be interpreted as a function.
By making certain assumptions,
amongst others one of real reflection c\oe fficients,
\cite{Ble87} shown that
\footnote{
A key point is that Bleistein showed that by virtue of stationary phase considerations
we can \textit{locally} consider
$
|\nabla g_k(\mathbf{r})|
\Leftrightarrow
2\cos\big( \theta_{inc}^{(i)}(\mathbf{r}_s,\mathbf{r}) \big)/c_{inc}(\mathbf{r})
$.
This implies a proper choice for $g_k$ that has the dimension of a time.
The reflectivity $\hat{R}$ then has the dimension of a time divided by a squared distance.
}
\begin{eqnarray}
\hat{R}_{inv}(\mathbf{r}_s,\mathbf{r})
&\approx&
\sum_{k\ge 1}
R(\mathbf{r},\theta_{inc}^{(i)}(\mathbf{r}_s,\mathbf{r}))
\Big(
\frac{2\cos\big( \theta_{inc}^{(i)}(\mathbf{r}_s,\mathbf{r}) \big)}{c_{inc}(\mathbf{r})}
\Big)^2
\delta_{bl}(g_k(\mathbf{r}))
\nonumber\\
\delta_{bl}(g_k(\mathbf{r}))
&=&
\frac{1}{\pi}\Re e\int
d\omega\hspace{0.5mm} e^{i\omega g_k(\mathbf{r})}F(\omega)
,
\label{eq:band_lim_delta}
\end{eqnarray}
where $F$ represents the residual (band-limited) wavelet present in the data (after pre-processing).
It maps in $\hat{R}_{inv}(\mathbf{r}_s,\mathbf{r})$ (also called reflectivity by slight abuse of language) through eq. (\ref{eq:band_lim_delta}).
We see that the band-limited Dirac $\delta_{bl}$
peaks where $g_k(\mathbf{r})=0$, i.e. on the reflector positions $S_k$.
This is why the inverted reflectivity $\hat{R}_{inv}$
is also called an image of the reflectors of the subsurface, or seismic image \cite{Tar05,Ble87,Cla85}.
We can demonstrate that $F$ maps in the image
in the direction perpendicular to the reflectors,
see Appendix \ref{app:Kirr1_3}.
Of course enough frequency and receiver aperture ranges are needed so that the inversion (\ref{eq:Kir_inv_1}) gives an unambiguous result,
depending amongst others on the number of samples that describe the reflectivity.

Eq. (\ref{eq:band_lim_delta}) allows us to deduce
(using $\forall \mathbf{r}\in S_k: g_k(\mathbf{r})=0$) \cite{Ble87}
\begin{eqnarray}
\forall \mathbf{r}\in S_k:
\quad
\hat{R}_{inv}(\mathbf{r}_s,\mathbf{r})
&\approx&
\alpha R(\mathbf{r},\theta_{inc}^{(i)}(\mathbf{r}_s,\mathbf{r}))
\Big(
\frac{2\cos\big( \theta_{inc}^{(i)}(\mathbf{r}_s,\mathbf{r}) \big)}{c_{inc}(\mathbf{r})}
\Big)^2
\nonumber\\
\alpha 
&=&
\frac{1}{\pi}\Re e\int
d\omega\hspace{0.5mm} F(\omega)
.
\label{eq:band_lim_delta2}
\end{eqnarray}
Suppose we could pick the amplitude variations along the amplitude peaks of continuous events
in the image, i.e. along reflector positions $S_k$,
and compute the incident angles $\theta_{inc}$ using rays \cite{Cer,Kra90} or 
wavefield decomposition techniques
and picked reflector dips.
Then, using the definition of the reflection c\oe fficient,
eq. (\ref{eq:ref72}), we can invert eq. (\ref{eq:band_lim_delta2}) for $c$
around reflector positions.
This common method of interpretation of the seismic image is called
``amplitude versus angle" (AVA) analysis  \cite{Rus88,Ble87}.

Now we can clarify what we previously meant by reflectors in a not-too-dense
configuration or separable almost everywhere.
From the Kirchhoff inversion point of view this means
reflectors separated almost everywhere from each other by more
than the source or receiver wavefield wavelengths at the dominant frequency.

\subsection{Questions.}
\label{sec:open}

We have recalled the approximations underlying Kirchhoff modelling and inversion.
We clarified some aspects concerning possibly non-smooth propagating media
and concerning the linearity approximation on reflectors
(Appendix \ref{app:Kirr1_main_steps}).
We now explore the following questions:
\begin{itemize}
\item
In the case of a very dense configuration of reflectors where it is no longer possible to separate each reflector
almost everywhere, can some reflectivity still be defined?
\item
Is it possible to model effects that go beyond first-order reflections (like first-order diffractions) within the Kirchhoff scheme, i.e. through a reflectivity?
\item
Can a general formula be set up for the conversion from reflectivity to velocity perturbation, and conversely?
This would offer opportunities on FWI approaches that include a reflectivity \cite{Xu12} or least-squares migration approaches based on Kirchhoff or Born modelling \cite{Sym16,Kro14}.
\item
In the traditional Kirchhoff inversion or true amplitude migration context, is it theoretically possible to interpret by AVA
more information in the image than only that associated with reflectors,
for instance the amplitudes of first-order diffractors?
\end{itemize}

\section{
Generalization of the Kirchhoff reflectivity.
}
\label{sec:KB}

\subsection{Born modelling approximation.}
\label{sec:Born_2}

We briefly recall the steps leading to Born modelling approximation \cite{Ble}.
We consider eq. (\ref{eq:wave1}) and decompose the subsurface velocity $c$ into
\begin{eqnarray}
\frac{1}{c^2(\mathbf{r})}
=
\frac{1}{c_0^2(\mathbf{r})}
+
\delta l(\mathbf{r})
,
\label{eq:Born_vel}
\end{eqnarray}
where $c_0$ is called a ``reference" medium velocity,
and $\delta l$ is the squared slowness related to
a ``perturbation" of the reference medium.
We decompose the subsurface wavefield into
$
P(\mathbf{r}_s,\mathbf{r},\omega)
=
P_0(\mathbf{r}_s,\mathbf{r},\omega)
+
P_{\delta l}(\mathbf{r}_s,\mathbf{r},\omega)
,
$
where the reference medium wavefield $P_0$ satisfies (in the time domain)
a scalar wave equation like eq. (\ref{eq:wave1}) in medium $c_0$ with source wavelet $s(t)$,
and $P_{\delta l}$ is the remaining wavefield (related to the perturbation
$\delta l$).
$P_{\delta l}$ can be decomposed into a linear contribution
and a non-linear contribution \cite{Ble}
$
P_{\delta l}(\mathbf{r}_s,\mathbf{r}_r,\omega)
=
P_{L}(\mathbf{r}_s,\mathbf{r}_r,\omega)
+
P_{NL}[P_{\delta l}](\mathbf{r}_s,\mathbf{r}_r,\omega)
,
$
where (considering the earth's surface at $z=0$)
\begin{eqnarray}
\label{eq:born_Plin}
P_{L}(\mathbf{r}_s,\mathbf{r}_r,\omega)
&=&
-(i \omega)^2
S(\omega)
\int_{z\ge 0}
d\mathbf{r} \hspace{0.5mm}
\delta l(\mathbf{r})
G_{0}(\mathbf{r}_s,\mathbf{r},\omega)
G_{0}(\mathbf{r},\mathbf{r}_r,\omega)
\label{eq:Born_1}\\
P_{NL}
[P_{\delta l}]
(\mathbf{r}_s,\mathbf{r}_r,\omega)
&=&
-(i \omega)^2
\int_{z\ge 0}
d\mathbf{r} \hspace{0.5mm}
\delta l(\mathbf{r})
P_{\delta l}(\mathbf{r}_s,\mathbf{r},\omega)
G_{0}(\mathbf{r},\mathbf{r}_r,\omega)
.
\nonumber
\end{eqnarray}
$G_0$ denotes the causal Green function in the reference medium,
i.e. satisfying eq. (\ref{eq:wave1})
with $c\rightarrow c_0$ and $s(t)=\delta(t)$.
Eq. (\ref{eq:Born_1}) does not involve any approximation.
Any reference medium can be chosen, in particular ones with non-smooth $c_0$,
thus also any ``strength" for the perturbation $\delta l$.

The Born modelling approximation
deals only with the linear contribution $P_{L}$
for the modelling of the wavefield perturbation
$$
P_{\delta l}(\mathbf{r}_s,\mathbf{r},\omega)\approx P_{L}(\mathbf{r}_s,\mathbf{r},\omega)
,
$$
which represents a good approximation if
\begin{eqnarray}
|P_{L}(\mathbf{r}_s,\mathbf{r},\omega)|>>|P_{NL}[P_{\delta l}](\mathbf{r}_s,\mathbf{r},\omega)|
\quad
\Leftarrow
\quad
\int_{z\ge 0}d\mathbf{r} \hspace{0.5mm}\frac{1}{c_0^2(\mathbf{r})}
>>
\Big|
\int_{z\ge 0}
d\mathbf{r} \hspace{0.5mm}\delta l(\mathbf{r})
\Big|
\quad
\text{and}
\quad
\frac{1}{c_0^2(\mathbf{r})}
>>
|\delta l(\mathbf{r})|
.
\label{eq:Born_app}
\end{eqnarray}
%
This implies that $c_0$ remains close to $c$ on average,
i.e. that $c_0$ reproduces the travel-times of first-order scattered events.

\subsection{Reformulation of Born modelling and generalized reflectivity.}
\label{sec:Born_3}

We reformulate the Born modelling equation
in a way that allows a direct comparison with the Kirchhoff modelling equation (\ref{eq:Kir_1_ref_final}).
Demonstrations that share a similar spirit can be found in \cite{Urs97,Ale02}.
Here we propose a different demonstration that involves travel-time branches,
and provide exhaustive detail on properties and consequences.

Firstly, we must constrain the perturbation $\delta l$ to describe
at least all reflectors of the true subsurface model $c$;
more generally, we constrain it to describe all discontinuities of the subsurface.
As a consequence $c_0$ will be smooth.
Secondly, as the Kirchhoff approximation involves 0-g.o. approximation
as discussed in \S \ref{sec:Kirr1},
we introduce 0-g.o. for the propagation of $G_0$.
As $c_0$ is smooth, we consider only direct (refracted) travel-time branches
\begin{eqnarray}
G_{0}(\mathbf{r}_s,\mathbf{r},\omega) 
&\approx&
\sum_{j=1}^{N(\mathbf{r}_s)}
G_{0}^{(j)}(\mathbf{r}_s,\mathbf{r},\omega) 
\nonumber\\
G_{0}^{(j)}(\mathbf{r}_s,\mathbf{r},\omega) 
&=&
A^{(j)}(\mathbf{r}_s,\mathbf{r}) e^{-i\omega T^{(j)}(\mathbf{r}_s,\mathbf{r})}
.
\label{eq:Cler_0go-2_2}
\end{eqnarray}
We use in this section similar notations as in \S \ref{sec:Kirr2_1_3}.
%
In the following we wish to be comparable to eq. (\ref{eq:Kir_1_ref_final})
and thus consider a single travel-time branch related to a smooth $c_0$
\footnote{
For other choices of $c_0$ and $\delta l$ where $c_0$ is non-smooth,
travel-time branches with reflections would have to be considered in $G_0$,
in the spirit Appendix \ref{app:Kirr1_main_steps}.
For instance if $\delta l$ describes only one reflector ($k$) present in $c$,
$c_0$ would contain the discontinuities related to all the other reflectors.
We do not detail this here.
}. 

We naturally constrain $c_0$ so that it reproduces the travel-times of the first-order events,
or in other words, so that it minimizes the travel-time corrections present in $P_{NL}$.
A velocity that meets this criterion as much as possible
can be defined through tomography \cite{Woo08,Lam08}.
There is thus a close link between $c_0$ and the smooth velocity $c_{inc}$
introduced in Kirchhoff modelling in \S \ref{sec:Kirr1}.
So, in the following we consider
\begin{eqnarray}
c_0(\mathbf{r})
\Leftrightarrow
c_{inc}(\mathbf{r})
\quad\text{and}\quad
G_{0}^{(j)}(\mathbf{r}_s,\mathbf{r},\omega) 
\Leftrightarrow
G_{inc}^{(j)}(\mathbf{r}_s,\mathbf{r},\omega) 
.
\label{eq:c_0_inc}
\end{eqnarray}

\begin{figure}[ht]
\begin{center}
\includegraphics[angle=0,width=0.7\linewidth]{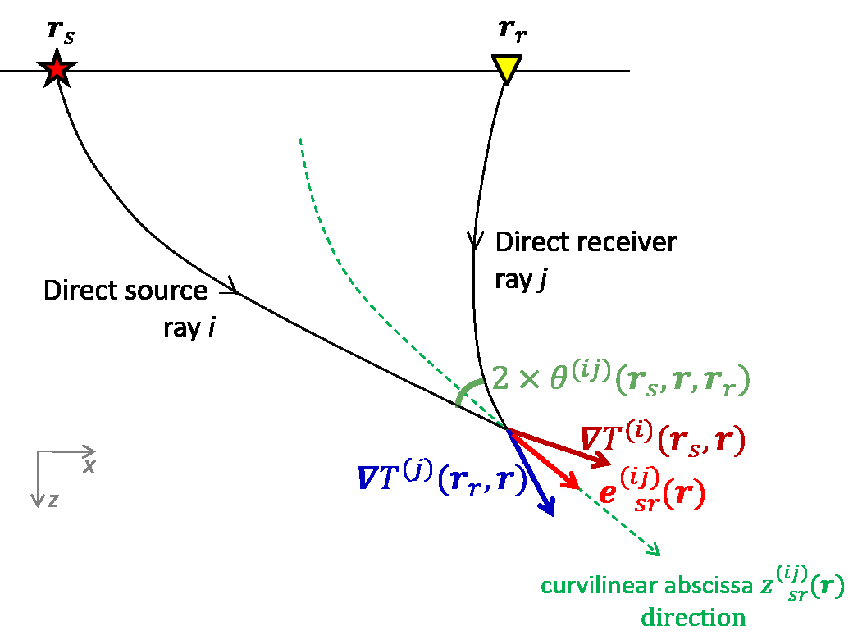}
\caption{
$\theta^{(ij)}$, $\mathbf{e}_{sr}^{(ij)}$ and $z_{sr}^{(ij)}$ representation for a source and a receiver direct travel-time branches in a smooth $c_0$.
\label{fig:zsr}
}
\end{center}
\end{figure}

We next denote by $\mathbf{r}_{sr}(\mathbf{r})=(x_{sr}(\mathbf{r}),y_{sr}(\mathbf{r}),z_{sr}(\mathbf{r}))$
any set of curvilinear coordinates obtained by transformation of the Cartesian coordinates $\mathbf{r}=(x,y,z)$;
the superscript ``$sr$" denotes that the curvilinear coordinates can be different for different
$\mathbf{r}_s$ and $\mathbf{r}_r$ positions.
The transformation must be well-defined, i.e. its Jacobian determinant must be non-null at every position $\mathbf{r}$.
To that aim, the curvilinear abscissa must not cross.
We choose a transformation
$\mathbf{r}_{sr}^{(ij)}(\mathbf{r})=(x_{sr}^{(ij)}(\mathbf{r}),y_{sr}^{(ij)}(\mathbf{r}),z_{sr}^{(ij)}(\mathbf{r}))$,
where $z_{sr}^{(ij)}(\mathbf{r})$ is a curvilinear coordinate in the average direction of a direct ray that links $\mathbf{r}_s$ to $\mathbf{r}$,
and of a direct ray that links $\mathbf{r}_r$ to $\mathbf{r}$, 
and $(x_{sr}^{(ij)}(\mathbf{r}),y_{sr}^{(ij)}(\mathbf{r}))$ are cartesian coordinates measured in the $(x,y)$ system,
see Fig.\ref{fig:zsr}.
The unit vector in the direction of the curvilinear abscissa $z_{sr}^{(ij)}(\mathbf{r})$ is
(using standard rules of 0-g.o., see Fig \ref{fig:zsr})
\begin{eqnarray}
&&
\mathbf{e}_{sr}^{(ij)}(\mathbf{r})
=
\frac{
\nabla
\big(
T^{(i)}(\mathbf{r}_s,\mathbf{r})+T^{(j)}(\mathbf{r}_r,\mathbf{r})
\big)
}{
\big|
\nabla
\big(
T^{(i)}(\mathbf{r}_s,\mathbf{r})+T^{(j)}(\mathbf{r}_r,\mathbf{r})
\big)
\big|
}
\quad
\Rightarrow
\quad
\frac{\partial}{\partial z_{sr}^{(ij)}(\mathbf{r})}
=
\mathbf{e}_{sr}^{(ij)}(\mathbf{r}).\nabla
\nonumber\\
&&
\frac{\partial}{\partial z_{sr}^{(ij)}(\mathbf{r})}
\big(
T^{(i)}(\mathbf{r}_s,\mathbf{r})+T^{(j)}(\mathbf{r}_r,\mathbf{r})
\big)
=
\big|
\nabla
\big(
T^{(i)}(\mathbf{r}_s,\mathbf{r})+T^{(j)}(\mathbf{r}_r,\mathbf{r})
\big)
\big|
=
\frac{
2\cos\big(
\theta^{(ij)}(\mathbf{r}_s,\mathbf{r},\mathbf{r}_r)
\big)
}{
c_{inc}(\mathbf{r})
}
.
\label{eq:esr2}
\end{eqnarray}
The Jacobian matrix of the transformation is invertible
because we have a unique incidence angle $\theta^{(ij)}$ at each position.
Using eq. (\ref{eq:Cler_0go-2_2}), we can show that for sufficiently large frequencies
(i.e. for the high-frequency leading term) we have
(the derivatives in the following are defined from the distributional derivative point of view)
\begin{eqnarray}
\label{eq:B_reformul_4_00}
&&
\forall \mathbf{r} \text{ such that }
\theta^{(ij)}(\mathbf{r}_s,\mathbf{r},\mathbf{r}_r)
\ne \pi/2:
\\
&&\quad\quad\quad
G_{inc}^{(i)}(\mathbf{r}_s,\mathbf{r},\omega)
G_{inc}^{(j)}(\mathbf{r},\mathbf{r}_r,\omega)
\approx
-\frac{1}{i\omega}
\frac{\partial}{\partial z_{sr}^{(ij)}(\mathbf{r})}
\Big\{
\frac{
c_{inc}(\mathbf{r})
}{
2\cos\big(
\theta^{(ij)}(\mathbf{r}_s,\mathbf{r},\mathbf{r}_r)
\big)
}
G_{inc}^{(i)}(\mathbf{r}_s,\mathbf{r},\omega)
G_{inc}^{(j)}(\mathbf{r},\mathbf{r}_r,\omega)
\Big\}
.
\nonumber
\end{eqnarray}
The area where 
$\theta^{(ij)}(\mathbf{r}_s,\mathbf{r},\mathbf{r}_r)=\pi/2$
corresponds to the area where one ray can be directly traced between $\mathbf{r}_s$ and $\mathbf{r}_r$,
i.e. to the ``diving waves" area.
To avoid singularities throughout this area, we constrain the Green functions in eq. (\ref{eq:Cler_0go-2_2})
to contain no diving-wave travel-time branches.
In areas where only diving waves occur in the subsurface, the Green functions $G_{inc}^{(i)}$ are thus null.
This does not reduce the generality of our considerations as,
using notations of \S \ref{sec:Born_2}, the diving waves
are described by the $P_0$ term
while the first-order scattered events
(first-order reflections and diffractions)
are described by the $P_{L}$ term.
We also constrain without loss of generality $\delta l$ to be null at the earth's surface for practical purposes.

We then insert eq. (\ref{eq:B_reformul_4_00}) in eq. (\ref{eq:Born_1}) and
use integration by parts
\begin{eqnarray}
\label{eq:B_reformul_5}
&&
\hspace{-0.9cm}
P_{L}(\mathbf{r}_s,\mathbf{r}_r,\omega)
\\
&\approx&
i \omega S(\omega)
\sum_{i=1}^{N(\mathbf{r}_s)}
\sum_{j=1}^{N(\mathbf{r}_r)}
\int_{z\ge 0}
d\mathbf{r} \hspace{0.5mm}
\delta l(\mathbf{r})
\frac{\partial}{\partial z_{sr}^{(ij)}(\mathbf{r})}
\Big\{
\frac{
c_{inc}(\mathbf{r})
}{
2\cos\big(
\theta^{(ij)}(\mathbf{r}_s,\mathbf{r},\mathbf{r}_r)
\big)
}G^{(i)}_{inc}(\mathbf{r}_s,\mathbf{r},\omega)
G^{(j)}_{inc}(\mathbf{r},\mathbf{r}_r,\omega)
\Big\}
\nonumber\\
&\approx&
i\omega S(\omega)
\sum_{i=1}^{N(\mathbf{r}_s)}
\sum_{j=1}^{N(\mathbf{r}_r)}
\int_{z\ge 0}
d\mathbf{r} \hspace{0.5mm}
\Big\{
-\frac{\partial\delta l(\mathbf{r})}{\partial z_{sr}^{(ij)}(\mathbf{r})}
\frac{
c_{inc}(\mathbf{r})
}{
2\cos\big(
\theta^{(ij)}(\mathbf{r}_s,\mathbf{r},\mathbf{r}_r)
\big)
}
\Big\}
G^{(i)}_{inc}(\mathbf{r}_s,\mathbf{r},\omega)
G^{(j)}_{inc}(\mathbf{r},\mathbf{r}_r,\omega)
\nonumber\\
&&
+
i\omega S(\omega)
\sum_{i=1}^{N(\mathbf{r}_s)}
\sum_{j=1}^{N(\mathbf{r}_r)}
\int_{z\ge 0} d\mathbf{r} \hspace{0.5mm}
\frac{\partial}{\partial z_{sr}^{(ij)}(\mathbf{r})}
\Big\{
\delta l(\mathbf{r})
\frac{
c_{inc}(\mathbf{r})
}{
2\cos\big(
\theta^{(ij)}(\mathbf{r}_s,\mathbf{r},\mathbf{r}_r)
\big)
}
G^{(i)}_{inc}(\mathbf{r}_s,\mathbf{r},\omega)
G^{(j)}_{inc}(\mathbf{r},\mathbf{r}_r,\omega)
\Big\}
\nonumber\\
&\approx&
i\omega S(\omega)
\sum_{i=1}^{N(\mathbf{r}_s)}
\sum_{j=1}^{N(\mathbf{r}_r)}
\int_{z\ge 0}
d\mathbf{r} \hspace{0.5mm}
\Big\{
-\frac{\partial\delta l(\mathbf{r})}{\partial z_{sr}^{(ij)}(\mathbf{r})}
\frac{
c_{inc}(\mathbf{r})
}{
2\cos\big(
\theta^{(ij)}(\mathbf{r}_s,\mathbf{r},\mathbf{r}_r)
\big)
}
\Big\}
G^{(i)}_{inc}(\mathbf{r}_s,\mathbf{r},\omega)
G^{(j)}_{inc}(\mathbf{r},\mathbf{r}_r,\omega)
.
\nonumber
\end{eqnarray}
Details leading to the last relationship are given in this footnote
\footnote{
We denote by $J^{(ij)}$ the absolute value of the Jacobian determinant of the transformation
$\mathbf{r}_{sr}^{(ij)}(\mathbf{r})=(x_{sr}^{(ij)}(\mathbf{r}),y_{sr}^{(ij)}(\mathbf{r}),z_{sr}^{(ij)}(\mathbf{r}))$ $\rightarrow$ $\mathbf{r}=(x,y,z)$.
We have (with slight abuses of notation, using 0-g.o., keeping the
high-frequency leading term in the penultimate line,
and using the Sommerfeld radiation
condition for $G^{}_{0}$ and that $\delta l$ is null at the earth's surface in the last line)
\begin{eqnarray}
&&
\int_{z\ge 0} d\mathbf{r} \hspace{0.5mm}
\frac{\partial}{\partial z_{sr}^{(ij)}(\mathbf{r})}
\Big\{
\delta l(\mathbf{r})
\frac{
c_{inc}(\mathbf{r})
}
{
2\cos\big(
\theta^{(ij)}(\mathbf{r}_s,\mathbf{r},\mathbf{r}_r)
\big)
}
G^{(i)}_{inc}(\mathbf{r}_s,\mathbf{r},\omega)
G^{(j)}_{inc}(\mathbf{r},\mathbf{r}_r,\omega)
\Big\}
\nonumber\\
&&
=
\int_{z_{sr}^{(ij)}\ge 0} d\mathbf{r}_{sr}^{(ij)} \hspace{0.5mm}
J^{(ij)}
\frac{\partial}{\partial z_{sr}^{(ij)}}
\Big\{
\delta l(\mathbf{r}_{sr}^{(ij)})
\frac{
c_{inc}(\mathbf{r}_{sr}^{(ij)})
}
{
2\cos\big(
\theta^{(ij)}(\mathbf{r}_s,\mathbf{r}_{sr}^{(ij)},\mathbf{r}_r)
\big)
}
G^{(i)}_{inc}(\mathbf{r}_s,\mathbf{r}_{sr}^{(ij)},\omega)
G^{(j)}_{inc}(\mathbf{r}_{sr}^{(ij)},\mathbf{r}_r,\omega)
\Big\}
\nonumber\\
&&
\approx
\int_{z_{sr}^{(ij)}\ge 0} d\mathbf{r}_{sr}^{(ij)} \hspace{0.5mm}
\frac{\partial}{\partial z_{sr}^{(ij)}}
\Big\{
J^{(ij)}
\delta l(\mathbf{r}_{sr}^{(ij)})
\frac{
c_{inc}(\mathbf{r}_{sr}^{(ij)})
}
{
2\cos\big(
\theta^{(ij)}(\mathbf{r}_s,\mathbf{r}_{sr}^{(ij)},\mathbf{r}_r)
\big)
}
G^{(i)}_{inc}(\mathbf{r}_s,\mathbf{r}_{sr}^{(ij)},\omega)
G^{(j)}_{inc}(\mathbf{r}_{sr}^{(ij)},\mathbf{r}_r,\omega)
\Big\}
\nonumber\\
&&
\approx
\int dx_{sr}^{(ij)}dy_{sr}^{(ij)}\hspace{0.5mm}
\Big[
J^{(ij)}
\delta l(\mathbf{r}_{sr}^{(ij)})
\frac{
c_{inc}(\mathbf{r}_{sr}^{(ij)})
}
{
2\cos\big(
\theta^{(ij)}(\mathbf{r}_s,\mathbf{r}_{sr}^{(ij)},\mathbf{r}_r)
\big)
}
G^{(i)}_{inc}(\mathbf{r}_s,\mathbf{r}_{sr}^{(ij)},\omega)
G^{(j)}_{inc}(\mathbf{r}_{sr}^{(ij)},\mathbf{r}_r,\omega)
\Big]_{z_{sr}^{(ij)}=0}^{z_{sr}^{(ij)}=+\infty}
\nonumber\\
&&
\approx
0
.
\nonumber
\end{eqnarray}
}.
This starts to look like the Kirchhoff modelling equation.
Again,
$\frac{\partial\delta l(\mathbf{r})}{\partial z_{sr}^{(ij)}(\mathbf{r})}$ in eq. (\ref{eq:B_reformul_5}) 
(where $\delta l$ contains discontinuities) is defined from the distributional derivative point of view.

We rewrite eq. (\ref{eq:B_reformul_5}) as
\begin{eqnarray}
\fbox{$
\begin{array}{rcl}
\label{eq:B_reformul_10_0}
&&
P_{L}(\mathbf{r}_s,\mathbf{r}_r,\omega)
\approx
\sum_{i=1}^{N(\mathbf{r}_s)}
\sum_{j=1}^{N(\mathbf{r}_r)}
\int_{z\ge 0}
d\mathbf{r} \hspace{0.5mm}
\hat{R}_{gen}(\mathbf{r},\theta^{(ij)}(\mathbf{r}_s,\mathbf{r},\mathbf{r}_r))
L_0^{(ij)}(\mathbf{r}_s,\mathbf{r}_r,\mathbf{r},\omega)
\\
&&
L_0^{(ij)}(\mathbf{r}_s,\mathbf{r}_r,\mathbf{r},\omega)
=
i\omega S(\omega)
G^{(i)}_{inc}(\mathbf{r}_s,\mathbf{r},\omega)
G^{(j)}_{inc}(\mathbf{r},\mathbf{r}_r,\omega)
\\
&&
\hat{R}_{gen}(\mathbf{r},\theta^{(ij)}(\mathbf{r}_s,\mathbf{r},\mathbf{r}_r))
=
-\frac{
c_{inc}(\mathbf{r})
}
{
2\cos\big(
\theta^{(ij)}(\mathbf{r}_s,\mathbf{r},\mathbf{r}_r)
\big)
}
\mathbf{e}_{sr}^{(ij)}(\mathbf{r}).\nabla \delta l(\mathbf{r})
.
\end{array}
$}
\end{eqnarray}

This main result consists of a reformulation of the Born approximation using 0-g.o. \cite{Urs97,Ale02}.
It looks like the Kirchhoff modelling equation (\ref{eq:Kir_1_ref_final}),
where $\hat{R}_{gen}$ is the counterpart of the Kirchhoff reflectivity distribution.
We call it ``generalized reflectivity".

We now have to understand precisely the differences between $\hat{R}_{gen}$
and Kirchhoff reflectivity $\hat{R}$.
Note that the Kirchhoff approximation describes first-order reflections,
whereas the Born approximation may contain more:
it can describe any first-order events present in $P_{L}$
(using notations of \S \ref{sec:Born_2})
like first-order diffractions.
We call $\hat{R}_{gen}$ ``generalized reflectivity"
even if it can describe more than reflections.

\subsection{Link between Born generalized reflectivity and Kirchhoff reflectivity.}
\label{sec:K_from_B_gR}

Kirchhoff modelling considers only reflectors in the subsurface.
We here firstly verify if Born generalized reflectivity $\hat{R}_{gen}$, eq. (\ref{eq:B_reformul_10_0}),
reduces to the Kirchhoff reflectivity $\hat{R}$, eq. (\ref{eq:Kir_1_ref_final}),
when the perturbation contains only reflectors.

We introduce the velocity perturbation $\delta c$ defined by
$$
c=c_{inc}+\delta c
.
$$
We have (using eqs. (\ref{eq:Born_vel}) and (\ref{eq:c_0_inc}))
$\delta l=1/(c_{inc}+\delta c)^2-1/c_{inc}^2$.
Because $\delta c$ must be sufficiently small,
we can perform a 1$^{st}$-order Taylor development and obtain
$$
\delta l(\mathbf{r})
\approx
-2\frac{\delta c(\mathbf{r})}{c_{inc}^3(\mathbf{r})}
.
$$
As $c_{inc}$ is smooth and $\delta c$ contains all the rapid velocity variations
of the subsurface, we can consider
$$
\nabla \delta l(\mathbf{r})
\approx
-\frac{2}{c_{inc}^3(\mathbf{r})}\nabla \delta c(\mathbf{r})
.
$$
Inserting this result in eq. (\ref{eq:B_reformul_10_0}) we obtain
$\hat{R}_{gen}$ as a function of the velocity perturbation
\begin{eqnarray}
\fbox{$
\begin{array}{rcl}
\label{eq:B_reformul_10}
\text{($c_{inc}$ smooth)}
\quad
\hat{R}_{gen}(\mathbf{r},\theta^{(ij)}(\mathbf{r}_s,\mathbf{r},\mathbf{r}_r))
=
\frac{1}
{
\cos\big(\theta^{(ij)}(\mathbf{r}_s,\mathbf{r},\mathbf{r}_r)\big)c_{inc}^2(\mathbf{r})
}
\mathbf{e}_{sr}^{(ij)}(\mathbf{r}).\nabla \delta c(\mathbf{r})
.
\end{array}
$}
\end{eqnarray}

Let us study what happens to the Born generalized reflectivity for a subsurface,
i.e. a perturbation $\delta c$, composed only of reflectors.
This can be modelled by
\begin{eqnarray}
\label{eq:dc}
&&
\delta c(\mathbf{r})
=
\sum_{k\ge 1}a_k(\mathbf{r})
\big[
H(g_k(\mathbf{r}))-0.5
\big]
,
\end{eqnarray}
where $H$ is the Heaviside function, and $a_k(\mathbf{r})$ a smooth
(continuously differentiable) function with compact support that has the dimension
of velocity, and simply ``adjusts" the Heaviside jumps.
When $\mathbf{r}$ is on reflector $k$, $a_k(\mathbf{r})$ equals the velocity jump $\Delta c(\mathbf{r})$ 
across the reflector (we use notation of eq. (\ref{eq:ref72}) where $\mathbf{n}$ is the normal to the reflectors continuously extended between reflectors)
\begin{eqnarray}
\label{eq:dc2}
&&
\forall \mathbf{r}\in S_k:
\quad
a_k(\mathbf{r})
=
\Delta c(\mathbf{r})
\nonumber\\
&&
\Delta c(\mathbf{r}) =
c_{+}(\mathbf{r})
-
c_{-}(\mathbf{r})
.
\end{eqnarray}
Evaluating the reflection c\oe fficient,
eq. (\ref{eq:ref72}) of Appendix \ref{app:Kirr1_1}, when
$\Delta c$ is small we obtain
the linearized reflection c\oe fficient $R_{lin}$ (to 1$^{st}$-order in $\delta c$)
%
%
\begin{eqnarray}
\begin{array}{rcl}
\label{eq:ref72'}
&&
R(\mathbf{r},\theta^{(ij)}(\mathbf{r}_s,\mathbf{r},\mathbf{r}_r))
\xrightarrow{\text{small $\delta c$}}
R_{lin}(\mathbf{r},\theta^{(ij)}(\mathbf{r}_s,\mathbf{r},\mathbf{r}_r))
=
\frac{\Delta c(\mathbf{r})}
{
2 c_{inc}(\mathbf{r}) \cos^2\big(\theta^{(ij)}(\mathbf{r}_s,\mathbf{r},\mathbf{r}_r)
\big)
}
.
\end{array}
\end{eqnarray}

We wish to compute the generalized reflectivity (\ref{eq:B_reformul_10})
corresponding to the velocity perturbation (\ref{eq:dc}).
We start by examining the $\mathbf{e}_{sr}^{(ij)}.\nabla\delta c$ term.
We have
\begin{equation}
\begin{array}{lll}
\mathbf{e}_{sr}^{(ij)}(\mathbf{r}).\nabla\delta c(\mathbf{r})
-
\sum_{k\ge 1} \mathbf{e}_{sr}^{(ij)}(\mathbf{r}).\nabla a_k(\mathbf{r}) 
\big[
H(g_k(\mathbf{r}))-0.5
\big]
&=
\sum_{k\ge 1} a_k(\mathbf{r}) \mathbf{e}_{sr}^{(ij)}(\mathbf{r}).\nabla H(g_k(\mathbf{r}))
&
\nonumber\\
&=
\sum_{k\ge 1}a_k(\mathbf{r}) \mathbf{e}_{sr}^{(ij)}(\mathbf{r}).\nabla g_k(\mathbf{r}) \frac{\partial}{\partial g_k} H(g_k(\mathbf{r}))
&
\nonumber\\
&=
\sum_{k\ge 1}a_k(\mathbf{r}) \mathbf{e}_{sr}^{(ij)}(\mathbf{r}).\nabla g_k(\mathbf{r}) \delta(g_k(\mathbf{r}))
&
\nonumber\\
&=
\sum_{k\ge 1}\Delta c(\mathbf{r}) \mathbf{e}_{sr}^{(ij)}(\mathbf{r}).\nabla g_k(\mathbf{r}) \delta(g_k(\mathbf{r}))
&
.
\nonumber
\end{array}
\end{equation}
When inserted into the modelling integral, eqs. (\ref{eq:B_reformul_10_0}) and (\ref{eq:B_reformul_10}),
the contribution of the
$\sum_{k\ge 1} \mathbf{e}_{sr}^{(ij)}(\mathbf{r}).\nabla a_k(\mathbf{r})\big[H(g_k(\mathbf{r}))-0.5\big]$ term
is negligible compared to the contribution of the
$\sum_{k\ge 1}\Delta c(\mathbf{r}) \mathbf{e}_{sr}^{(ij)}(\mathbf{r}).\nabla g_k(\mathbf{r}) \delta(g_k(\mathbf{r}))$ term.
Indeed, the second term leads to a sum of surface integrals (because of $\delta(g_k(\mathbf{r}))$)
for which stationary phases exist for reflections whereas
the first term (where $a_k$ is smooth) leads to a sum of volume integrals
\footnote{
Each $a_k$ has a compact support around reflector $k$ positions.
The ``second term" has the same compact support.
Nevertheless this leads to a volume integral.
}
for which stationary phases exist for diving waves only.
As the Green functions $G^{(i)}_{inc}$ entering into our Born modelling equation were constrained to contain no diving wave, the impact of the second term can be neglected in our case.
The dominant contribution of $\mathbf{e}_{sr}^{(ij)}(\mathbf{r}).\nabla\delta c(\mathbf{r})$ is thus given by
(using eq. (\ref{eq:ref72'}))
\begin{eqnarray}
\mathbf{e}_{sr}^{(ij)}(\mathbf{r}).\nabla\delta c(\mathbf{r})
&\rightarrow&
\sum_{k\ge 1}\Delta c(\mathbf{r}) \mathbf{e}_{sr}^{(ij)}(\mathbf{r}).\nabla g_k(\mathbf{r}) \delta(g_k(\mathbf{r}))
\nonumber\\
&=&
\sum_{k\ge 1}
R_{lin}(\mathbf{r},\theta^{(ij)}(\mathbf{r}_s,\mathbf{r},\mathbf{r}_r))2 c_{inc}(\mathbf{r})
\cos^2\big(\theta^{(ij)}(\mathbf{r}_s,\mathbf{r},\mathbf{r}_r)\big)
\mathbf{e}_{sr}^{(ij)}(\mathbf{r}).\nabla g_k(\mathbf{r}) \delta(g_k(\mathbf{r}))
.
\nonumber
\end{eqnarray}
%
%
%
Inserting this result in eq. (\ref{eq:B_reformul_10}) gives
\begin{eqnarray}
\hat{R}_{gen}(\mathbf{r},\theta^{(ij)}(\mathbf{r}_s,\mathbf{r},\mathbf{r}_r))
=
\sum_{k\ge 1} 
R_{lin}(\mathbf{r},\theta^{(ij)}(\mathbf{r}_s,\mathbf{r},\mathbf{r}_r))
\frac{
2\cos\big(\theta^{(ij)}(\mathbf{r}_s,\mathbf{r},\mathbf{r}_r)\big)
}
{
c_{inc}(\mathbf{r})
}
\mathbf{e}_{sr}^{(ij)}(\mathbf{r}).\nabla g_k(\mathbf{r})
\delta(g_k(\mathbf{r}))
.
\label{eq:Bf4}
\end{eqnarray}
We can consider, again from stationary phase reasoning \cite{Ble,Ble87,Urs97}
$$
\mathbf{e}_{sr}^{(ij)}(\mathbf{r}).\nabla g_k(\mathbf{r})
\Leftrightarrow
|\nabla g_k(\mathbf{r})|
.
$$
Finally, using eq. (\ref{eq:sing_ft}), we obtain
\begin{eqnarray}
\hat{R}_{gen}(\mathbf{r},\theta^{(ij)}(\mathbf{r}_s,\mathbf{r},\mathbf{r}_r))
=
\sum_{k\ge 1} 
R_{lin}(\mathbf{r},\theta^{(ij)}(\mathbf{r}_s,\mathbf{r},\mathbf{r}_r))
\frac{
2\cos\big(\theta^{(ij)}(\mathbf{r}_s,\mathbf{r},\mathbf{r}_r)\big)
}
{
c_{inc}(\mathbf{r})
}
\delta_{S_k}(\mathbf{r})
.
\end{eqnarray}
Gathering previous results we have 
\begin{eqnarray}
\fbox{$
\begin{array}{rcl}
\label{eq:Bf6}
&&
\text{\underline{Born generalized reflectivity:}}
\\
&&
\\
&&
\hat{R}_{gen}(\mathbf{r},\theta^{(ij)}(\mathbf{r}_s,\mathbf{r},\mathbf{r}_r))
\\
&&
\\
&&
\hspace{0.5cm}
=
\frac{1}
{
\cos\big(\theta^{(ij)}(\mathbf{r}_s,\mathbf{r},\mathbf{r}_r)\big)c_{inc}^2(\mathbf{r})
}
\mathbf{e}_{sr}^{(ij)}(\mathbf{r}).\nabla \delta c(\mathbf{r})
\\
&&
\hspace{1.5cm}
\text{$\rightarrow$ describes first-order effects related to any kind of small perturbations $\delta c$}
\\
&&
\hspace{2cm}
\text{(first-order reflections, diffractions)}
\\
&&
\\
&&
\hspace{0.5cm}
\xrightarrow{\text{reflectors only}}
\sum_{k\ge 1} 
R_{lin}(\mathbf{r},\theta^{(ij)}(\mathbf{r}_s,\mathbf{r},\mathbf{r}_r))
\frac{
2\cos\big(\theta^{(ij)}(\mathbf{r}_s,\mathbf{r},\mathbf{r}_r)\big)
}
{
c_{inc}(\mathbf{r})
}
\delta_{S_k}(\mathbf{r})
\\
&&
\hspace{8cm}
\text{with $\theta^{(ij)}(\mathbf{r}_s,\mathbf{r},\mathbf{r}_r)\Leftrightarrow\theta_{inc}^{(i)}(\mathbf{r}_s,\mathbf{r})$}
\\
&&
\hspace{1.5cm}
\text{$\rightarrow$ describes first-order reflections related to sufficiently weak velocity} 
\\
&&
\hspace{2cm}
\text{discontinuities.}
\end{array}
$}
\end{eqnarray}
%
For comparison let us remember the content of Kirchhoff reflectivity, eq. (\ref{eq:Kir_1_ref_final})
\begin{eqnarray}
\fbox{$
\begin{array}{rcl}
\label{eq:Kf6}
&&
\text{\underline{Kirchhoff reflectivity:}}
\\
&&
\\
&&
\hat{R}(\mathbf{r},\theta^{(ij)}(\mathbf{r}_s,\mathbf{r},\mathbf{r}_r))
\\
&&
\\
&&
\hspace{0.5cm}
=
\sum_{k\ge 1} 
R(\mathbf{r},\theta^{(ij)}(\mathbf{r}_s,\mathbf{r},\mathbf{r}_r))
\frac{
2\cos\big(\theta^{(ij)}(\mathbf{r}_s,\mathbf{r},\mathbf{r}_r)\big)
}
{
c_{inc}(\mathbf{r})
}
\delta_{S_k}(\mathbf{r})
\quad
\text{with $\theta^{(ij)}(\mathbf{r}_s,\mathbf{r},\mathbf{r}_r)\Leftrightarrow\theta_{inc}^{i}(\mathbf{r}_s,\mathbf{r})$}
\\
&&
\hspace{1.5cm}
\text{$\rightarrow$ describes first-order (possibly postcritical) reflections related to }
\\
&&
\hspace{2cm}
\text{ (possibly larger) velocity discontinuities.}
\end{array}
$}
\end{eqnarray}

Let us discuss what we have learned until now.
From the propagation point of view,
Born $c_0$ has been constrained here to be very close to Kirchhoff's $c_{inc}$
(we considered them as identical).
According to considerations of \S \ref{sec:Kirr1_1bis} and Appendix \ref{app:Kirr1_1},
Kirchhoff has a slight advantage
over Born because Kirchhoff modelling may be more ``effective"
(the general form of the linearity approximation on reflectors
allows different modellings for events above and below strong reflectors\dots).

From the reflectivity point of view,
we notice main differences if we compare the Born generalized reflectivity (\ref{eq:Bf6})
to Kirchhoff reflectivity (\ref{eq:Kf6}):
\begin{enumerate}
\item
For reflections:\\
Born is based on a linearized version $R_{lin}$
of the 0-g.o. reflection c\oe fficient $R$,
whereas Kirchhoff is based on the full 0-g.o. reflection c\oe fficient
$R$ defined by eq. (\ref{eq:ref72}).
This represents an advantage for Kirchhoff in describing the following:
\begin{enumerate}
\item
reflections due to larger velocity contrasts,
\item
larger incidence angle reflections,
\item
complex reflections.
\end{enumerate}
These effects affect the phase of the wavefield,
and are thus contained in the non-linear term $P_{NL}$
(using notations of \S \ref{sec:Born_2}).
Because the Born approximation does not account for it,
these effects cannot be included in Born generalized reflectivity.
As those effects that are non-linear with respect to
the wavefield perturbation are still linear
with respect to the full 0-g.o. reflection c\oe fficient,
they may be accounted for by Kirchhoff.
But remind that the linearity approximation on reflectors physically implies
not too large velocity contrasts, limiting the ranges of validity of points (a)-(c).
\item
For very dense reflector configurations:\\
The generalized reflectivity remains well defined in the limit of a
configuration where it is no longer possible to separate each reflector almost everywhere.
\item
For other first-order events (like first-order diffractions):\\
Born can describe them whereas traditional Kirchhoff cannot.
The price to pay is that the reflectivity must then
depend explicitly on the receiver positions $\mathbf{r}_r$ through
$\theta^{(ij)}(\mathbf{r}_s,\mathbf{r},\mathbf{r}_r)$
(obvious for diffractions because they radiate in every direction).
\end{enumerate}
According to the first point, Kirchhoff contains more than Born.
According to the second and third points, Born contains more than Kirchhoff.
The considerations of this article allow us to gather the strengths of both schemes in a unique scheme:
one may use Born's generalized reflectivity $\hat{R}_{gen}$ together with
the full reflection c\oe fficient $R$ instead of the linearized reflection c\oe fficient $R_{lin}$
to model the reflections (only).


To close this section, we give few words about the application of the 3D Radon transform (RT) \cite{Mil87} to the generalized reflectivity.
We show in Appendix \ref{app:RT} that the right-hand side term of the generalized reflectivity (\ref{eq:Bf6}) can be decomposed into
\begin{eqnarray}
\label{eq:dc4bis}
\mathbf{e}_{sr}^{(ij)}(\mathbf{r}).\nabla \delta c(\mathbf{r})
=
-\mathbf{e}_{sr}^{(ij)}(\mathbf{r}).\int d\mathbf{\xi} \hspace{0.3mm} \Delta \hat{c}(  \mathbf{\xi}, \mathbf{\xi}.\mathbf{r})\mathbf{\xi}
\quad\text{with}\quad
\Delta \hat{c}(\mathbf{\xi},p)
=
\frac{1}{8\pi^2}
\frac{\partial^3}{\partial p^3}
\delta \hat{c}(  \mathbf{\xi}, p )
,
\end{eqnarray}
where $\delta \hat{c}$ represents the RT of $\delta c$
defined by
\begin{eqnarray}
\delta\hat{c}( \mathbf{\xi}, p )
=
\int d\mathbf{r} 
\delta(p- \mathbf{\xi} . \mathbf{r} )
\delta c(\mathbf{r})
.
\label{eq:dc3bis}
\end{eqnarray}
$\mathbf{\xi}$ is a dimensionless unit vector
and $p$ the distance from the origin of the plane perpendicular to $\mathbf{\xi}$ that goes through $\mathbf{r}$.
$d\mathbf{\xi}$ denotes the solid angle
measure over the unit sphere surrounding $\mathbf{r}$.
%
Eq. (\ref{eq:dc4bis}) can be interpreted as a decomposition of $\nabla \delta c(\mathbf{r})$ into a continuous set of planes, or planar "interfaces" 
with "slopes" $\mathbf{\xi}$.
$\Delta \hat{c}(  \mathbf{\xi}, \mathbf{\xi}.\mathbf{r})$ represents the weight 
of each "slope" in $\nabla \delta c(\mathbf{r})$.
Using eq. (\ref{eq:dc4bis}),
we see that the slopes couple to $\mathbf{e}^{sr}(\mathbf{r})$ through $\mathbf{e}^{sr}(\mathbf{r}).\xi$ 
%
\footnote{
Interestingly, there is a similarity between the RT continuous decomposition of $\delta c$ over planar "virtual interfaces"
and the discrete decomposition of $\delta c$ over physical reflectors related to eqs. (\ref{eq:dc}) and (\ref{eq:dc2}), see Appendix \ref{app:RT}.
}.

The decomposition (\ref{eq:dc4bis}) together with eq. (\ref{eq:Bf6}) has the formal interest to
define how the generalized reflectivity can be decomposed into local slopes, 
and to give information on the slopes that contribute the most to the modelling
(thus to the stationary phase)
through $\Delta \hat{c}(  \mathbf{\xi}, \mathbf{\xi}.\mathbf{r})$.
($\Delta \hat{c}(  \mathbf{\xi}, \mathbf{\xi}.\mathbf{r})$ will tend to be constant in every direction $\mathbf{\xi}$ for a diffractor spike in $\mathbf{r}$,
whereas it will tend to be much larger in one direction for a reflector\dots)
Also, this decomposition may have the practical interest to avoid a direct computation of $\nabla \delta c(\mathbf{r})$, that could be very noisy
for some rapid variations of $\delta c(\mathbf{r})$ (for instance diffractor spikes).

\section{Some theoretical considerations on the utility of the generalized reflectivity for interpretation.}
\label{sec:inv}

The Kirchhoff inversion procedure introduced in \S \ref{sec:Kirr2_1_3} has more flexibility than
the direct inversion of the Born modelling equation.
Indeed, the latter attempts to directly recover a material property of the subsurface $\delta l$
(or $\delta c$), which is independent of the source position.
The least-squares inversion of Born modelling equation (\ref{eq:Born_1}) is a procedure
that combines the data from all sources
($P$ represents data recorded at the earth's surface, pre-processed to retain only first-order events)
\begin{eqnarray}
\min_{\delta l(\mathbf{r})}
\int
d\mathbf{r}_s \hspace{0.5mm}
\int
d\omega 
\int
d\mathbf{r}_r \hspace{0.5mm}
\Big|
P(\mathbf{r}_s,\mathbf{r}_r,\omega)
+
(i \omega)^2
S(\omega)
\int_{z\ge 0}
d\mathbf{r} \hspace{0.5mm}
\delta l(\mathbf{r})
G_{0}(\mathbf{r}_s,\mathbf{r},\omega)
G_{0}(\mathbf{r},\mathbf{r}_r,\omega)
\Big|^2
.
\label{eq:Born_inv_ls}
\end{eqnarray}
Kirchhoff inversion attempts to recover a reflectivity (or a seismic image)
that is not a material property of the subsurface and depends on the source position.
Thus Kirchhoff least-squares inversion, eq. (\ref{eq:Kir_inv_1}),
is done for each shot (or offset if the data are reorganized) independently.
As a second step, the material properties of the subsurface are recovered
by inverting the obtained reflectivity.
This step has already been presented in \S \ref{sec:Kirr2_1_3}
to interpret amplitudes of the reflectors present in the seismic image.
In the following we show how the generalized reflectivity concept allows us
to interpret events other
than reflectors in the seismic image.

Suppose we recorded data $P$ at the earth's surface and pre-processed
to retain only first-order events,
with non-linear (or higher-order) events filtered out (in particular multiple reflections).
Suppose we also produced a smooth subsurface model $c_{inc}$ that allows us to compute $G_{inc}^{(i)}$ and $\theta^{(ij)}$ for chosen direct travel-time branches.
Traditional Kirchhoff inversion, eq. (\ref{eq:Kir_inv_1}), inverts for a reflectivity
$\hat{R}_{inv}(\mathbf{r}_s,\mathbf{r})$ that d\oe s not depend on the receiver positions $\mathbf{r}_r$
\cite{Ble87}.

As the generalized reflectivity considerations
fit into a Kirchhoff modelling framework, eq. (\ref{eq:Bf6})
can directly be used to interpret the result of a Kirchhoff inversion (\ref{eq:Kir_inv_1}).
Or course the most exhaustive scheme to invert for the generalized reflectivity
should depend on the receiver
positions because $\theta^{(ij)}(\mathbf{r}_s,\mathbf{r},\mathbf{r}_r)$ is included in the definition of the generalized reflectivity.
This would not cause any formal difficulties but would lead to an inversion scheme that is different
from the commonly implemented scheme that does not depend on the receiver positions, eq. (\ref{eq:Kir_inv_1}),
with additional ill-posed problems.
Here we want our considerations to be applicable to the commonly implemented scheme.
The obtained $\hat{R}_{inv}$ then represents a band-limited average of $\hat{R}_{gen}$,
eq. (\ref{eq:Bf6}), over the receiver positions.
If $S_{rec}$ denotes the area over which the receiver are spread, we have
\begin{eqnarray}
\fbox{$
\begin{array}{rcl}
\label{eq:kir_stand_av}
\hat{R}_{inv}(\mathbf{r}_s,\mathbf{r})
&\approx&
\frac{1}{S_{rec}}\int_{S_{rec}} d\mathbf{r}_r\hspace{1mm}\hat{R}_{gen}^{bl}(\mathbf{r},\theta^{(ij)}(\mathbf{r}_s,\mathbf{r},\mathbf{r}_r))
\\
&&
\\
&\approx&
\mathbf{a}(\mathbf{r}_s,\mathbf{r})
.\nabla \delta c_{bl}(\mathbf{r})
\quad
\text{where}
\quad
\mathbf{a}(\mathbf{r}_s,\mathbf{r})
=
\frac{1}{c_{inc}^2(\mathbf{r})}
\frac{1}{S_{rec}}
\int_{S_{rec}} d\mathbf{r}_r\hspace{1mm}
\frac{
1
}
{
\cos
\big(
\theta^{(ij)}(\mathbf{r}_s,\mathbf{r},\mathbf{r}_r)
\big)
}
\mathbf{e}_{sr}^{(ij)}(\mathbf{r})
,
\end{array}
$}
\end{eqnarray}
where $\delta c_{bl}$ is the band-limited version of the velocity perturbation
(because in practice we deal with band-limited Kirchhoff inversion).
Inverting eq. (\ref{eq:kir_stand_av}) for each shot would allow us to convert the seismic image (i.e. the reflectivity $\hat{R}_{inv}$ computed by traditional Kirchhoff inversion schemes) into a velocity perturbation $\delta c_{bl}$ and vice-versa.
More generally one could also invert eq. (\ref{eq:Bf6}) for each source-receiver configuration.
This offers opportunities for further interpretation of seismic images and also for
FWI approaches that include a reflectivity,
showing how to rigorously convert the reflectivity into a velocity perturbation.

We discussed in \S \ref{sec:Kirr2_1_3} the traditional way to interpret seismic images,
considering only the amplitudes of reflectors.
But seismic data and images also contain events that are not related to reflections,
for instance those related to diffractions.
It has been kinematically understood why migration
collapses first-order diffractions in the seismic image \cite{Cla85,Aki80},
and the generalized reflectivity considerations explain this from a fundamental point of view.
It also opens some doors:
eq. (\ref{eq:kir_stand_av})
can be used to interpret events related to first-order diffractions,
i.e. recover the perturbation $\delta c_{bl}$ that generates them.

If we deal with with a stack over shots, we have from eq. (\ref{eq:kir_stand_av})
\begin{eqnarray}
\label{eq:a_stack}
\int d\mathbf{r}_s \hspace{0.5mm}
\hat{R}_{inv}(\mathbf{r}_s,\mathbf{r})
&\approx&
\alpha(\mathbf{r})\hspace{1mm}\mathbf{e}_{average}(\mathbf{r})
.\nabla \delta c_{bl}(\mathbf{r})
\\
\mathbf{e}_{average}(\mathbf{r})
&=&
\frac{1}{\alpha(\mathbf{r})}\mathbf{a}_{stack}(\mathbf{r})
\quad\text{with}\quad
\alpha(\mathbf{r})=
\sqrt{\mathbf{a}_{stack}(\mathbf{r}).\mathbf{a}_{stack}(\mathbf{r})}
\nonumber\\
\mathbf{a}_{stack}(\mathbf{r})
&=&
\frac{1}{c_{inc}^2(\mathbf{r})}
\frac{1}{S_{rec}}
\int_{S_{rec}} d\mathbf{r}_r\hspace{1mm}
\int d\mathbf{r}_s \hspace{0.5mm}
\frac{
1
}
{
\cos
\big(
\theta^{(ij)}(\mathbf{r}_s,\mathbf{r},\mathbf{r}_r)
\big)
}
\mathbf{e}_{sr}^{(ij)}(\mathbf{r})
.
\nonumber
\end{eqnarray}
The direction of the unit vector $\mathbf{e}_{average}$ depends on the acquisition configuration
and the propagation in the subsurface (through the "illumination direction" $\mathbf{e}_{sr}^{(ij)}$ and $\theta^{(ij)}$).
As already mentioned in \S \ref{sec:Kirr2_1_3} this behavior is different from reflectors,
for which a residual wavelet maps in the direction perpendicular to the reflectors whatever the acquisition.
For spike diffractors, the wavelet will map in the $\mathbf{e}_{average}$ direction in the image stacked over shots,
this direction depending on the acquisition configuration.

We denote by $\eta(\mathbf{r})$ the curvilinear abscissa defined by
$\mathbf{e}_{average}(\mathbf{r})$
and consider the transformation
$\mathbf{r}=(x,y,z)\rightarrow(x_s(\mathbf{r}),y_s(\mathbf{r}),\eta(\mathbf{r}))$,
where $x_s(\mathbf{r})$ and $y_s(\mathbf{r})$ define the positions at the earth surface associated to $\eta(\mathbf{r})$.
We have
\begin{eqnarray}
\label{eq:e_av}
\mathbf{e}_{average}(\mathbf{r})
.\nabla
=
\frac{\partial}{\partial \eta(\mathbf{r})}
.
\end{eqnarray}
From eqs. (\ref{eq:a_stack}) and (\ref{eq:e_av}), one has to integrate
$\frac{1}{\alpha(\mathbf{r})}\int d\mathbf{r}_s \hspace{0.5mm}
\hat{R}_{inv}(\mathbf{r}_s,\mathbf{r})
$ over the curvilinear abscissa $\eta$ to recover the band-limited velocity perturbation $\delta c_{bl}$.
We obtain, with slight abuse of notation
\begin{eqnarray}
\fbox{$
\begin{array}{rcl}
\delta c_{bl}(\mathbf{r})
=
\int_{0}^{\eta(\mathbf{r})}
d\eta'\hspace{1mm}
\frac{1}{\alpha(x_s(\mathbf{r}),y_s(\mathbf{r}),\eta')}
\int d\mathbf{r}_s \hspace{0.5mm}
\hat{R}_{inv}(\mathbf{r}_s;x_s(\mathbf{r}),y_s(\mathbf{r}),\eta')
+
constant
.
\label{eq:R_oth_mig-2_int}
\end{array}
$}
\end{eqnarray}
$\mathbf{e}_{average}$, eq. (\ref{eq:a_stack}), is computable
through knowledge of the propagation directions of the source and receiver wavefields.
The $constant$ can be computed by imposing that $\delta c_{bl}(\mathbf{r})$ is null at the earth's surface.
Thus the interpretation of more information
than that associated with reflectors is theoretically possible by generalized reflectivity considerations.

\section{Conclusions.}
\label{sec:concl0}

We recalled the chain of approximations leading to Kirchhoff and Born modelling equations.
They both contain a linearization but each offers some specifics.
The Kirchhoff approximation allows, for example, the modelling of first-order reflections on stronger
discontinuities and postcritical reflections.
Born approximation makes it possible to model reflections on weak discontinuities only,
but also first-order events beyond reflections
(like first-order diffractions). 
We pointed out from a fundamental point of view the strengths and weaknesses of these schemes.

We took the opportunity to clarify some aspects related to Kirchhoff modelling approximation,
concerning possibly non-smooth propagating media
and the linearity approximation on reflectors.
We discussed how Kirchhoff and Born modelling lead a general expression
for the conversion from velocity model perturbation to
reflectivity (and conversely) through the generalized reflectivity concept.

The generalized reflectivity offers opportunities that have been discussed formally in the article:
\begin{itemize}
\item
On FWI approaches that include a reflectivity or least squares migration approaches that can be based on Kirchhoff or Born modelling:
to rigorously convert the reflectivity into a velocity perturbation.
\item
In the framework of traditional Kirchhoff inversion or true amplitude migration,
i.e. for the interpretation of seismic-migrated images:
to interpret by AVA (amplitude versus angle)
more information than the amplitudes associated to first-order reflections,
for instance the amplitudes of first-order diffractors.
Also, it would theoretically allow us to go beyond AVA analysis,
inverting for the whole seismic image amplitude information (not only amplitude information at peaks)
to recover the related velocity model perturbation.
\item
In the framework of traditional Kirchhoff modelling scheme:
to model first-order effects that go beyond first-order reflections (like first-order diffractions).
\end{itemize}


\section*{Acknowledgments.}

The author is particularly indebted to Gilles Lambar\'e, Samuel Gray
and Jean Virieux for numerous enlightening
discussions and profound recommendations.
The author is grateful to CGG for the permission to publish this work,
and to Thibaut Allemand and Aramide Moronfoye for proofreading of the manuscript.


\begin{appendix}

\section{Chain of approximations leading to the Kirchhoff modelling approximation equation.}
\label{app:Kirr1_main_steps}

We remind the steps that lead to eq. (\ref{eq:Kir6-kbisbis}).

\subsection{Kirchhoff modelling considering one reflector.}
\label{app:Kirr1_1}

Considering the notations of \S \ref{sec:Kirr1} and using the representation theorem \cite{Aki80,Ble},
some manipulations allow us to demonstrate
the so-called ``Kirchhoff integral''
(that does not involve any approximation so far)
%
\begin{eqnarray}
G_{ref}(\mathbf{r}_s,\mathbf{r}_r,\omega)
=
\int_{S_k} d\mathbf{r} \hspace{0.5mm} 
\big(
\nabla G_{ref}(\mathbf{r}_s,\mathbf{r},\omega) G_{inc}(\mathbf{r},\mathbf{r}_r,\omega)
-
G_{ref}(\mathbf{r}_s,\mathbf{r},\omega) \nabla G_{inc}(\mathbf{r},\mathbf{r}_r,\omega)
\big)
.\mathbf{n}(\mathbf{r})
.
\label{eq:repres2}
\end{eqnarray}
%
$G_{inc}$ is known from eq. (\ref{eq:wave1}) solved in the medium above the reflector $S_k$ (with $s(t)=\delta(t)$), i.e. with radiation (or ``absorbing" boundary) conditions on  $S_k$.

$G_{ref}$, the Green function that describes the events generated on $S_k$ and below, is obtained by solving eq. (\ref{eq:repres2}).
The latter is not an easy task because the integral
depends on the $G_{ref}$ and $\mathbf{n}.\nabla G_{ref}$ values on  $S_k$.
The Kirchhoff approximation allows us to ease this task by
finding approximate $G_{ref}$ and $\mathbf{n}.\nabla G_{ref}$ values for those that enter into the integral.

We consider the 0-g.o. approximation, eq. (\ref{eq:Cler_0go-2bis}),
where the Green functions $G_{inc}^{(i)}$ are related to each of the
travel-time branches (or ray paths) in the incident medium.
The essence of the Kirchhoff approximation is to assume the following relationship
between the Green functions $G_{inc}^{(i)}$ and $G_{ref}^{(i)}$ along $S_k$:
\begin{eqnarray}
\forall \mathbf{r}\in S_k:
\quad\quad\quad\quad
G_{ref}^{(i)}(\mathbf{r}_s,\mathbf{r},\omega)
&\approx&
R^{(i)}(\mathbf{r}_s,\mathbf{r})
G_{inc}^{(i)}(\mathbf{r}_s,\mathbf{r},\omega)
\nonumber\\
\nabla G_{ref}^{(i)}(\mathbf{r}_s,\mathbf{r},\omega).\mathbf{n}(\mathbf{r})
&\approx&
- R^{(i)}(\mathbf{r}_s,\mathbf{r})
\nabla G_{inc}^{(i)}(\mathbf{r}_s,\mathbf{r},\omega)
.\mathbf{n}(\mathbf{r})
,
\label{eq:Kir3}
\end{eqnarray}
where $R^{(i)}$ is a possibly complex function called the reflection coefficient defined in eq. (\ref{eq:ref72}).
Eq. (\ref{eq:Kir3}) is justified within 0-g.o. approximation when $S_k$ is a reflector;
it implies that we consider single events reflected from above on $S_k$ and that
(for now) we neglect events generated below $S_k$.
(A reflector is also called a ``smooth interface" \cite{Ble} because its
surface must have sufficiently small curvature and be associated with a reflection c\oe fficient that
varies slowly enough along the interface.)

Inserting eq. (\ref{eq:Kir3}) in eq. (\ref{eq:repres2}), using the 0-g.o. approximation (\ref{eq:Cler_0go-2bis}) for $G_{inc}^{(i)}$ and keeping only the high-frequency leading terms gives \cite{Ble}
\begin{eqnarray}
&&
G_{ref}(\mathbf{r}_s,\mathbf{r}_r,\omega)
=
\sum_{i,j\ge0}
G_{ref}^{(ij)}(\mathbf{r}_s,\mathbf{r}_r,\omega)
\nonumber\\
&&
G_{ref}^{(ij)}(\mathbf{r}_s,\mathbf{r}_r,\omega)
=
i \omega
\int_{S_k}
d\mathbf{r} \hspace{0.5cm}
R(\mathbf{r},\theta_{inc}^{(i)}(\mathbf{r}_s,\mathbf{r}))\hspace{1mm}
\mathbf{n}(\mathbf{r}).\nabla \big( T^{(i)}(\mathbf{r}_s,\mathbf{r})+T^{(j)}(\mathbf{r}_r,\mathbf{r}) \big)
\nonumber\\
&&
\hspace{5.5cm}
\times
A^{(i)}(\mathbf{r}_s,\mathbf{r})
A^{(j)}(\mathbf{r}_r,\mathbf{r})
e^{-i\omega ( T^{(i)}(\mathbf{r}_s,\mathbf{r}) + T^{(j)}(\mathbf{r}_r,\mathbf{r}) )}
,
\label{eq:Kir6}
\end{eqnarray}
where $G_{ref}^{(ij)}$ contains one reflection event on $S_k$. 
It represents the contribution 
of the $i^{th}$ ``source" travel-time branch, coupled with a single reflection
from above on reflector $k$ to the $j^{th}$ ``receiver" travel-time branch.
Again, events generated below $S_k$ are neglected for now.

Eq. (\ref{eq:Kir6}) is not symmetric (reciprocal) under the exchange of $\mathbf{r}_s$ and $\mathbf{r}_r$
because of the angle $\theta_{inc}^{(i)}(\mathbf{r}_s,\mathbf{r})$
that depends on $\mathbf{r}_s$ and $i$ (and implicitly contains knowledge of the geological dip at every position on the reflector),
but does not depend on $\mathbf{r}_r$ and $j$.
This is conceptually annoying because Green functions should satisfy
symmetry under the exchange of $\mathbf{r}_s$ and $\mathbf{r}_r$ \cite{Ble}.
To recover this symmetry we use the property that the phase of eq. (\ref{eq:Kir6})
is stationary when the Snell-Descartes law for reflections is satisfied \cite{Ble,Urs97},
i.e. for ``specular" source and receiver ray pairs.
Then, using the stationary phase approximation principle described for instance in  \cite{Ble},
valid for sufficiently high frequencies,
we can make the following replacements in eq. (\ref{eq:Kir6}) \cite{Ble,Ble87,Urs97}:
\begin{eqnarray}
\label{eq:theta_replac}
\forall \mathbf{r}\in S_k:
\hspace{2.5cm}
\theta_{inc}^{(i)}(\mathbf{r}_s,\mathbf{r})
\quad
&\Leftrightarrow&
\quad
\theta^{(ij)}(\mathbf{r}_s,\mathbf{r},\mathbf{r}_r)
\\
R(\mathbf{r},\theta_{inc}^{(i)}(\mathbf{r}_s,\mathbf{r}))
\quad
&\Leftrightarrow&
\quad
R(\mathbf{r},\theta^{(ij)}(\mathbf{r}_s,\mathbf{r},\mathbf{r}_r))
\nonumber\\
\mathbf{n}(\mathbf{r}).\nabla \big( T^{(i)}(\mathbf{r}_s,\mathbf{r})+T^{(j)}(\mathbf{r}_r,\mathbf{r}) \big)
\quad
&\Leftrightarrow&
\quad
\big| \nabla \big( T^{(i)}(\mathbf{r}_s,\mathbf{r})+T^{(j)}(\mathbf{r}_r,\mathbf{r}) \big) \big|
=
2\cos\big( \theta^{(ij)}(\mathbf{r}_s,\mathbf{r},\mathbf{r}_r) \big)/c(\mathbf{r})
.
\nonumber
\end{eqnarray}
%
Note that the last line of eq. (\ref{eq:theta_replac}) is equivalent to the replacement
$\mathbf{n}(\mathbf{r})\Leftrightarrow\mathbf{e}^{(ij)}_{sr}(\mathbf{r})$ where $\mathbf{e}^{(ij)}_{sr}$ represents the average direction of source and receivers rays.
By inserting those results in eq. (\ref{eq:Kir6}), factorizing the 0-g.o.-Green functions,
and introducing the source signature to deal with wavefields, we obtain
\begin{eqnarray}
&&
P_{ref}(\mathbf{r}_s,\mathbf{r}_r,\omega)
=
\sum_{i,j\ge 0}
P_{ref}^{(ij)}(\mathbf{r}_s,\mathbf{r}_r,\omega)
\nonumber\\
&&
P_{ref}^{(ij)}(\mathbf{r}_s,\mathbf{r}_r,\omega)
=
\int_{S_k}
d\mathbf{r} \hspace{0.5mm}
R(\mathbf{r},\theta^{(ij)}(\mathbf{r}_s,\mathbf{r},\mathbf{r}_r))
\frac{2\cos\big( \theta^{(ij)}(\mathbf{r}_s,\mathbf{r},\mathbf{r}_r) \big)}
{c(\mathbf{r})}
L_{inc}^{(ij)}(\mathbf{r}_s,\mathbf{r}_r,\mathbf{r},\omega)
\nonumber\\
&&
L_{inc}^{(ij)}(\mathbf{r}_s,\mathbf{r}_r,\mathbf{r},\omega)
=
i \omega S(\omega) 
G_{inc}^{(i)}(\mathbf{r}_s,\mathbf{r},\omega)
G_{inc}^{(j)}(\mathbf{r},\mathbf{r}_r,\omega)
,
\label{eq:Kir6-2}
\end{eqnarray}
where $P_{ref}(\mathbf{r}_s,\mathbf{r}_r,\omega)=S(\omega)G_{ref}(\mathbf{r}_s,\mathbf{r}_r,\omega)$
denotes the total wavefield reflected on $S_k$ and measured at the earth's surface.
Each $P_{ref}^{(ij)}(\mathbf{r}_s,\mathbf{r}_r,\omega)=S(\omega)G_{ref}^{(ij)}(\mathbf{r}_s,\mathbf{r}_r,\omega)$ is related to a single reflection event on $S_k$
for the corresponding source and receiver travel-time branches reaching the reflector.
This is the so-called Kirchhoff modelling approximation equation for one reflector $S_k$.

Still within 0-g.o., several extensions of those approximations exist \cite{Cer,Kra90,Bra03,Kro98,Sto02}.
Here we keep the simplest one because it does not affect the conclusions of this article.
Also, one can write eq. (\ref{eq:Kir6-2}) in other domains than the source-receivers domain involving 
the Beylkin determinant \cite{Bey85,Bey86}.

%
%
%
%


\subsection{Details on the linearity approximation on reflectors.}
\label{app:Kirr1_2}

Until now we have considered events occurring on a single reflector
in Kirchhoff modelling equation (\ref{eq:Kir6-2}).
Now suppose the subsurface reflectors are in a
configuration where they are separable almost everywhere,
i.e. a not too dense configuration (in a sense clarified in \S  \ref{sec:Kirr2_1_3}).
Each reflector is identified by $k\in \mathbb{N}^*$.
The idea behind the linearity approximation on reflectors is to consider
a Kirchhoff modelling equation like eq. (\ref{eq:Kir6-2}) for each reflector and to sum them,
in order to account for the contributions of all reflectors.
This introduces additional physical limitations.

We add subscript $k$ in eq. (\ref{eq:Kir6-2}) to make explicit that it concerns reflector $k$:
obviously to $P_{ref,k}^{(ij)}$ and $P_{ref,k}^{(ij)}$,
but also to $\theta^{(ij)}_k$ and $G_{inc,k}^{(i)}$
because they describe results of propagation in the medium above $S_k$ excluding $S_k$,
thus different propagation results when different reflectors $k$ are considered.
We obtain
\begin{eqnarray}
&&
P_{ref}(\mathbf{r}_s,\mathbf{r}_r,\omega)
=
\sum_{k\ge 1}
\sum_{i,j\ge 1}
P_{ref,k}^{(ij)}(\mathbf{r}_s,\mathbf{r}_r,\omega)
\nonumber\\
&&
P_{ref,k}^{(ij)}(\mathbf{r}_s,\mathbf{r}_r,\omega)
=
\int_{S_k}
d\mathbf{r} \hspace{0.5mm}
R(\mathbf{r},\theta^{(ij)}_k(\mathbf{r}_s,\mathbf{r},\mathbf{r}_r))
\frac{2\cos\big( \theta^{(ij)}_k(\mathbf{r}_s,\mathbf{r},\mathbf{r}_r) \big)}
{c(\mathbf{r})}
L_{inc,k}^{(ij)}(\mathbf{r}_s,\mathbf{r}_r,\mathbf{r},\omega)
\nonumber\\
&&
L_{inc,k}^{(ij)}(\mathbf{r}_s,\mathbf{r}_r,\mathbf{r},\omega)
=
i \omega S(\omega) 
G_{inc,k}^{(i)}(\mathbf{r}_s,\mathbf{r},\omega)
G_{inc,k}^{(j)}(\mathbf{r},\mathbf{r}_r,\omega)
.
\label{eq:Kir6-k}
\end{eqnarray}
Again, in $P_{ref,k}^{(ij)}$, only single reflections from above on reflector $k$ are considered
through the integral,
even if multipathing (direct arrivals and multiple reflections above reflector $k$) is considered through the travel-time branches.
Note that crossing reflectors can naturally be considered in eq. (\ref{eq:Kir6-k})
including a $k$-dependency to the normal $\mathbf{n}(\mathbf{r})$, leading to an explicitly $k$-dependent reflection coefficient through eq. (\ref{eq:ref72}).


We discuss why the contributions of each reflector can be described separately and then summed.
The Born approximation (which is detailed in \S \ref{sec:Born_2}) tells us that
first-order scattering effects (such as first-order, or single, reflections and diffractions)
can be modelled linearly regarding the wavefield
if the velocity perturbation is not too strong.
Applied to Kirchhoff modelling (based on the reflection coefficient and not on velocity perturbations),
this linearity implies that each single reflection event recorded at the earth's surface $P_{ref,k}^{(ij)}$
can be associated with one Kirchhoff modelling equation of the form (\ref{eq:Kir6-2})
if the reflection coefficients are not too large.
The total reflected wavefield is obtained by ``summing" the $P_{ref,k}^{(ij)}$
(here over the reflectors and the travel-time branches).
Although single reflections on reflector $k$ are considered in each $P_{ref,k}^{(ij)}$,
reflection events can still be ``strong".
%
%
Thus, even if eq. (\ref{eq:Kir6-k}) is mathematically well defined for large contrasts,
it is physically valid only for not too large ones.

The result obtained here, eq. (\ref{eq:Kir6-k}), represents the traditional Kirchhoff modelling approximation scheme with multiple travel-time branches and different Green functions for different reflectors.
It makes it possible to model the reflected events part of the subsurface wavefield measured at the earth's surface.

In the $P_{ref,k}^{(ij)}$ of eq. (\ref{eq:Kir6-k}) we allow
possible multiple reflections on reflectors above $S_k$
during propagations from source or to receivers
(but, rigorously, not on $S_k$ even if this restriction can be overcome)
through the travel-time branches,
and single reflections from above on $S_k$ through the integral.
This allows us to keep the problem linear in terms of the reflection coefficients.
Note that an extension to multiple reflections from above and below on $S_k$
has been proposed, see e.g. \cite{Ber82}.
Then the problem becomes non-linear in terms of the reflection coefficients \cite{Weg97,Weg18,Kro02,Mal09}.

We now describe the further approximations that lead to the traditional Kirchhoff modelling scheme.
The traditional linearity approximation on reflectors \cite{Ble}
considers only the $P_{ref,k}^{(ij)}$ contributions to $P_{ref}$
related to direct (or refracted) source and receiver travel-time branches,
i.e. $i\in[1,N(\mathbf{r}_s)]$ and $j\in[1,N(\mathbf{r}_r)]$.
Consequently, single (first-order) reflections on  $S_k$ only are modelled
and $G_{inc,k}^{(i)}$ and $\theta_k^{(ij)}$ can be considered as independent of reflectors $k$
(the presence of reflectors above a subsurface position does not condition directly the number of direct waves
reaching the subsurface position).
We denote them by $G_{inc}^{(i)}$ and $\theta^{(ij)}$
and finally obtain eq. (\ref{eq:Kir6-kbisbis}).
%

\section{Time-depth equivalency.}
\label{app:Kirr1_3}

Time-depth equivalency is often mentioned in seismic.
It, among others things, implies that the residual wavelet $f(t)$ present in Kirchhoff-inverted reflectivity
maps in the direction perpendicular to the reflectors.
In other terms $\delta_{bl}(g_k(\mathbf{r}))$ in eq. (\ref{eq:band_lim_delta})
represents a band-limited delta distribution that ``peaks" in the direction perpendicular
to the interface $k$ \cite{Ble,Ble87}.
This is what we demonstrate now.

We perform a 1$^{st}$-order Taylor expansion of $g_k(\mathbf{r})$ around 
the orthogonal projection of $\mathbf{r}$ on interface $k$,
denoted by $\mathbf{r}_k(\mathbf{r})$.
We obtain
\begin{eqnarray}
g_k(\mathbf{r})
&=&
g_k\big(\mathbf{r}_k(\mathbf{r})\big)
+
\nabla_{\mathbf{r}_k} g_k (\mathbf{r}_k(\mathbf{r}))
.
\big(\mathbf{r}-\mathbf{r}_k(\mathbf{r})\big)
+
o(|\mathbf{r}-\mathbf{r}_k(\mathbf{r})|^2)
\nonumber\\
&=&
\frac{
\nabla_{\mathbf{r}_k} g_k (\mathbf{r}_k(\mathbf{r}))
}
{
|\nabla_{\mathbf{r}_k} g_k (\mathbf{r}_k(\mathbf{r}))|
}
.
\big(\mathbf{r}-\mathbf{r}_k(\mathbf{r})\big)
|\nabla_{\mathbf{r}_k} g_k (\mathbf{r}_k(\mathbf{r}))|
+
o(|\mathbf{r}-\mathbf{r}_k(\mathbf{r})|^2)
\nonumber\\
&=&
\mathbf{n}(\mathbf{r}_k(\mathbf{r}))
.
\big(\mathbf{r}-\mathbf{r}_k(\mathbf{r})\big)
|\nabla_{\mathbf{r}_k} g_k (\mathbf{r}_k(\mathbf{r}))|
+
o(|\mathbf{r}-\mathbf{r}_k(\mathbf{r})|^2)
,
\label{eq:below_7}
\end{eqnarray}
where
$\mathbf{n}(\mathbf{r}_k(\mathbf{r}))
=
\frac{\nabla_{\mathbf{r}_k} g_k (\mathbf{r}_k(\mathbf{r}))}{|\nabla_{\mathbf{r}_k} g_k (\mathbf{r}_k(\mathbf{r}))|}$
is the unit vector normal to the interface $k$ at position $\mathbf{r}_k(\mathbf{r})$ that points ``downward".
The remainder term is negligible within 0-g.o.,
i.e. when the interface curvature (the 2$^{nd}$-order derivatives of $g_k$) is sufficiently small.
Using eq. (\ref{eq:below_7}), we have
\begin{eqnarray}
\delta_{bl}(g_k(\mathbf{r}))
&\approx&
\frac{1}{|\nabla_{\mathbf{r}_k} g_k (\mathbf{r}_k(\mathbf{r}))|}
\delta_{bl}
\Big(
\mathbf{n}(\mathbf{r}_k(\mathbf{r}))
.
\big(\mathbf{r}-\mathbf{r}_k(\mathbf{r})\big)
\Big)
.
\label{eq:below_7_ds2}
\end{eqnarray}
This demonstrates that the smearing of $\delta_{bl}(g_k(\mathbf{r}))$ (due to the residual wavelet $F$)
occurs in the direction $\mathbf{n}(\mathbf{r}_k(\mathbf{r}))$ perpendicular to the reflector.

\section{Radon transform and Born velocity perturbation.}
\label{app:RT}

The Radon transform of a function $f(\mathbf{r})$ of a 3D variable is defined by \cite{Mil87}
\begin{equation}
\hat{f}( \mathbf{\xi}, p )
=
\int d\mathbf{r} 
\delta(p- \mathbf{\xi} . \mathbf{r} )
f(\mathbf{r})
,
\end{equation}
where $\mathbf{\xi}$ is the dimensionless unit vector perpendicular to a plane
and $p$ the distance from the plane to the origin.
$(\mathbf{\xi},p)$ defines the positions $\mathbf{r}$ on the plane through $p- \mathbf{\xi} . \mathbf{r}=0$.

The inverse Radon transform is given by \cite{Mil87}
\begin{equation}
f(\mathbf{r})
=
-\frac{1}{8\pi^2}
\int dp \hspace{0.3mm}
d\mathbf{\xi} \hspace{0.3mm}
\delta''(p- \mathbf{\xi} . \mathbf{r} )
f(  \mathbf{\xi}, p )
,
\end{equation}
which can be also written (using $\delta(p)=\partial H(p) / \partial p$)
\begin{equation}
f(\mathbf{r})
=
\frac{1}{8\pi^2}
\int dp \hspace{0.3mm}
d\mathbf{\xi} \hspace{0.3mm}
H(p- \mathbf{\xi} . \mathbf{r} )
\frac{\partial^3}{\partial p^3}
\hat{f}(  \mathbf{\xi}, p )
,
\label{eq:RT1}
\end{equation}
where $d\mathbf{\xi}$ denotes the solid angle
measure over the unit sphere surrounding $\mathbf{r}$.
Applying the 3D Radon transform to the Born velocity perturbation we obtain \cite{Mil87}
\begin{eqnarray}
&&
\delta c(\mathbf{r})
=
\int dp \hspace{0.3mm}
d\mathbf{\xi} \hspace{0.3mm}
\Delta \hat{c}(\mathbf{\xi},p)
H(p- \mathbf{\xi} . \mathbf{r} )
\nonumber\\
&&
\Delta \hat{c}(\mathbf{\xi},p)
=
\frac{1}{8\pi^2}
\frac{\partial^3}{\partial p^3}
\delta \hat{c}(  \mathbf{\xi}, p )
\quad\text{with}\quad
\delta\hat{c}( \mathbf{\xi}, p )
=
\int d\mathbf{r} 
\delta(p- \mathbf{\xi} . \mathbf{r} )
\delta c(\mathbf{r})
.
\label{eq:dc3}
\end{eqnarray}
$\delta c(\mathbf{r})$ is thus decomposed
into a summation over planes $(\mathbf{\xi}, p)$
associated to a weight $\Delta \hat{c}(\mathbf{\xi},p)$.
Interestingly, eq. (\ref{eq:dc3}) has a similar form
than eqs. (\ref{eq:dc}) and (\ref{eq:dc2})
but, instead of considering a discrete sum over true reflectors of the subsurface,
it considers a continuous sum over "virtual" planar reflectors
($g_k(\mathbf{r})$ is replaced by $p-\mathbf{\xi}.\mathbf{r}$ and $a_k(\mathbf{r})$ by $\Delta \hat{c}(\mathbf{\xi},p)$).

Using $\mathbf{\nabla}_{\mathbf{r}} H(p-\mathbf{\xi}.\mathbf{r})= -\mathbf{\xi}\hspace{0.3mm}\delta(p-\mathbf{\xi}.\mathbf{r})$, we obtain
\begin{eqnarray}
\label{eq:dc4}
\nabla \delta c(\mathbf{r})
=
-
\int d\mathbf{\xi} \hspace{0.3mm} \Delta \hat{c}(  \mathbf{\xi}, \mathbf{\xi}.\mathbf{r})\mathbf{\xi}
.
\end{eqnarray}

\end{appendix}

\bibliographystyle{apalike}
\bibliography{biblio}

\end{document}